\pdfoutput=1
\documentclass{article}
\usepackage[hscale=0.8,vscale=0.8]{geometry}

\usepackage{graphicx}
\usepackage{url}
\usepackage{fixltx2e}
\usepackage{amsmath}
\usepackage{multirow}
\usepackage{algorithm}
\usepackage[noend]{algpseudocode}
\usepackage[hang]{subfigure}
\usepackage{amsthm}
\usepackage{amssymb}
\usepackage{pifont}
\newcommand{\cmark}{\ding{51}}%
\newcommand{\xmark}{\ding{55}}%
\newtheorem{theorem}{Theorem}[section]
\newtheorem{lemma}{Lemma}[section]
\newtheorem{corollary}{Corollary}[section]
\newtheorem{definition}{Definition}[section]
\newcommand{\SetProblem}{{\sc $(W,\epsilon)$-Approximate Set Membership}}
\newcommand{\SetMulProblem}{{\sc $(W,\epsilon)$-Approximate Set Multiplicity}}
\newcommand{\CountDistinctProblem}{{\sc $(W,\epsilon,\delta)$-Approximate Count Distinct}}
\newcommand{\EntropyProblem}{{\sc $(W,\epsilon,\delta)$-Entropy Estimation}}
\renewcommand{\hat}{\widehat}

\ifdefined\NINEPAGES
\else
\newcommand*{\EXTENDED}{}
\fi

\ifdefined\NINEPAGES
\allowdisplaybreaks

\DeclareMathAlphabet{\mathpzc}{OT1}{pzc}{m}{it}
\fi
\begin{document}

\date{}
\author{
	Eran Assaf\\
	Hebrew University\\
	\and
	Ran Ben Basat\\
	Technion\\
	\and
	Gil Einziger\\
	Nokia Bell Labs\\
	\and
	Roy Friedman\\
	Technion\\
}
\title{Pay for a Sliding Bloom Filter and Get Counting, Distinct Elements, and Entropy for Free}



\maketitle

\renewcommand{\epsilon}{\varepsilon}
\newcommand{\entropy}{{\sc Entropy}}
\newcommand{\entropyVariable}{\ensuremath{\widehat{H}}}
\newcommand{\distinctLB}{{\sc DistinctLB}}
\newcommand{\distinctMLE}{{\sc DistinctMLE}}
\newcommand{\ecor}{$\epsilon$-correct}
\newcommand{\qsr}{\emph{$q$-Sliding Ranker}}
\newcommand{\last}{\ensuremath{\mathit{last}}}
\newcommand{\poly}{\mbox{poly}}
\newcommand{\ceil}[1]{ \left\lceil{#1}\right\rceil}
\newcommand{\floor}[1]{ \left\lfloor{#1}\right\rfloor}
\newcommand{\parentheses}[1]{ \left({#1}\right)}
\newcommand{\abs}[1]{ \left|{#1}\right|}
\newcommand{\logp}[1]{\log\parentheses{#1}}
\newcommand{\Omegap}[1]{\Omega\parentheses{#1}}
\newcommand{\Thetap}[1]{\Theta\parentheses{#1}}
\newcommand{\omegap}[1]{\omega\parentheses{#1}}
\newcommand{\logc}[1]{\log\ceil{#1}}
\newcommand{\flogp}[1]{\floor{\logp{#1}}}
\newcommand{\clogp}[1]{\ceil{\logp{#1}}}
\newcommand{\clogc}[1]{\ceil{\logc{#1}}}
\newcommand{\cdotpa}[1]{\cdot\parentheses{#1}}
\newcommand{\inc}[1]{$#1 = #1 + 1$}
\newcommand{\oneOverE}{ \eps^{-1} }
\newcommand{\range}[2][0]{#1,1,\ldots,#2}
\newcommand{\frange}[1]{\set{\range{#1}}}
\newcommand{\xrange}[1]{\frange{#1-1}}
\newcommand{\smallMultError}{(1+o(1))}
\newcommand{\brackets}[1]{\left[#1\right]}
\newcommand{\lowerbound}{\max \set{\log W ,\frac{1}{2\epsilon+W^{-1}}}}
\newcommand{\smallEpsLowerbound}{\window\logp{\frac{1}{\weps}}}
\newcommand{\smallEpsMemoryTheta}{$\Theta\parentheses{\smallEpsMemoryConsumption}$}
\newcommand{\smallEpsMemoryConsumption}{W\cdot\logp{\frac{1}{\weps}}}

\newcommand{\largeEpsRestriction}{For any \largeEps{},}
\newcommand{\largeEps}{\ensuremath{\eps^{-1} \le 2W\left(1-\frac{1}{\logw}\right)}}
\newcommand{\smallEpsRestriction}{For any \smallEps{},}
\newcommand{\smallEps}{$\eps^{-1}>2W\left(1-\frac{1}{\logw}\right)=2\window(1-o(1))$}
\newcommand{\bc}{{\sc Basic-Counting}}
\newcommand{\bs}{{\sc Basic-Summing}}
\newcommand{\windowcounting}{ {\sc $(W,\epsilon)$-Window-Counting}}

\newcommand{\query}[1][] { {\sc Query}$(#1)$}
\newcommand{\add}  [1][] { {\sc Add}$(#1)$}

\newcommand{\window}{W}
\newcommand{\logw}{\log \window}
\newcommand{\logrw}[1][]{\log^{#1}{\bsrange\window}}
\newcommand{\flogw}{\floor{\log \window}}
\newcommand{\weps}{\window\epsilon}
\newcommand{\logweps}{\log{\weps}}
\newcommand{\bitarray}{b}
\newcommand{\currentBlockIndex}{i}
\newcommand{\currentBlock}{\bitarray_{\currentBlockIndex}}
\newcommand{\remainder}{y}
\newcommand{\numBlocks}{k}
\newcommand{\sumOfBits}{B}
\newcommand{\blockSize}{\frac{\window}{\numBlocks}}
\newcommand{\iblockSize}{\frac{\numBlocks}{\window}}
\newcommand{\threshold}{\blockSize}
\newcommand{\halfBlock}{\frac{\window}{2\numBlocks}}
\newcommand{\blockOffset}{m}
\newcommand{\inputVariable}{x}

\newcommand{\bcTableColumnWidth}{1.5cm}
\newcommand{\bsTableColumnWidth}{1.7cm}
\newcommand{\bsExtendedTableColumnWidth}{3cm}
\newcommand{\bcExtendedTableColumnWidth}{2.8cm}
\newcommand{\bcNarrowTableColumnWidth}{1.5cm}
\newcommand{\bsNarrowTableColumnWidth}{1.5cm}
\newcommand{\bsWorstCaseTableColumnWidth}{2cm}

\newcommand{\bsrange}{ R }
\newcommand{\bsReminderPercisionParameter}{ \gamma }
\newcommand{\bsest}{ \widehat{\bssum}}
\newcommand{\bssum}{ S^W }
\newcommand{\bsFracInput}{ \inputVariable' }
\newcommand{\bserror}{ \bsrange\window\epsilon }
\newcommand{\bsfractionbits}{ \frac{\bsReminderPercisionParameter}{\epsilon} }
\newcommand{\bsReminderFractionBits}{ \upsilon}
\newcommand{\bsAnalysisTarget}{ \bssum}
\newcommand{\bsAnalysisEstimator}{ \widehat{\bsAnalysisTarget}}
\newcommand{\bsAnalysisError}{ \bsAnalysisEstimator - \bsAnalysisTarget}
\newcommand{\bsRoundingError}{ \xi}

\begin{abstract}
For many networking applications, recent data is more significant than older data, motivating the need for sliding window solutions.
Various capabilities, such as DDoS detection and load balancing, require insights about multiple metrics including Bloom filters, per-flow counting, count distinct and entropy estimation.

In this work, we present a unified construction that solves all the above problems in the sliding window model.
Our single solution offers a better space to accuracy tradeoff than the state-of-the-art for each of these individual problems!
We show this both analytically and by running multiple real Internet backbone and datacenter packet traces.
\end{abstract}

\section{Introduction}
Network measurements are at the core of many applications,
such as load balancing, quality of service, anomaly/intrusion detection, and caching~\cite{CONGA,DevoFlow,TinyLFU,IntrusionDetection2,ApproximateFairness}. Measurement algorithms are required to cope with the throughput demands of modern links,
forcing them to rely on scarcely available fast SRAM memory.
However, such memory is limited in size~\cite{CounterBraids}, which motivates approximate solutions that conserve space.

Network algorithms often find recent data useful. For example, anomaly detection systems attempt to detect manifesting anomalies and a load balancer needs to balance the current load rather than the historical one.
Hence, the sliding window model is an active research~field~\cite{WCSS,slidngBloomFilterInfocom,Naor2013,ActiveActive,TBF}.

The desired measurement types differ from one application to the other. For example, a load balancer may be interested in the heavy hitter flows~\cite{CONGA}, which are responsible for a large portion of the traffic. Additionally, anomaly detection systems often monitor the number of distinct elements~\cite{IntrusionDetection2} and entropy~\cite{Entropy1} or use Bloom filters~\cite{AnomalyBF}.
Yet, existing algorithms usually provide just a single utility at a time, e.g., approximate set membership (Bloom filters)~\cite{Bloom}, per-flow counting~\cite{SpectralBloom,TinyTable}, count distinct~\cite{CD3,CD0,CD4} and entropy~\cite{Entropy1}.
Therefore, as networks complexity grows, multiple measurement types may be required.
However, employing multiple stand-alone solutions incurs the additive combined cost of each of them, which is inefficient in both memory and computation.

\begin{figure}[t]
	\center{
		\includegraphics[width = .9\columnwidth]{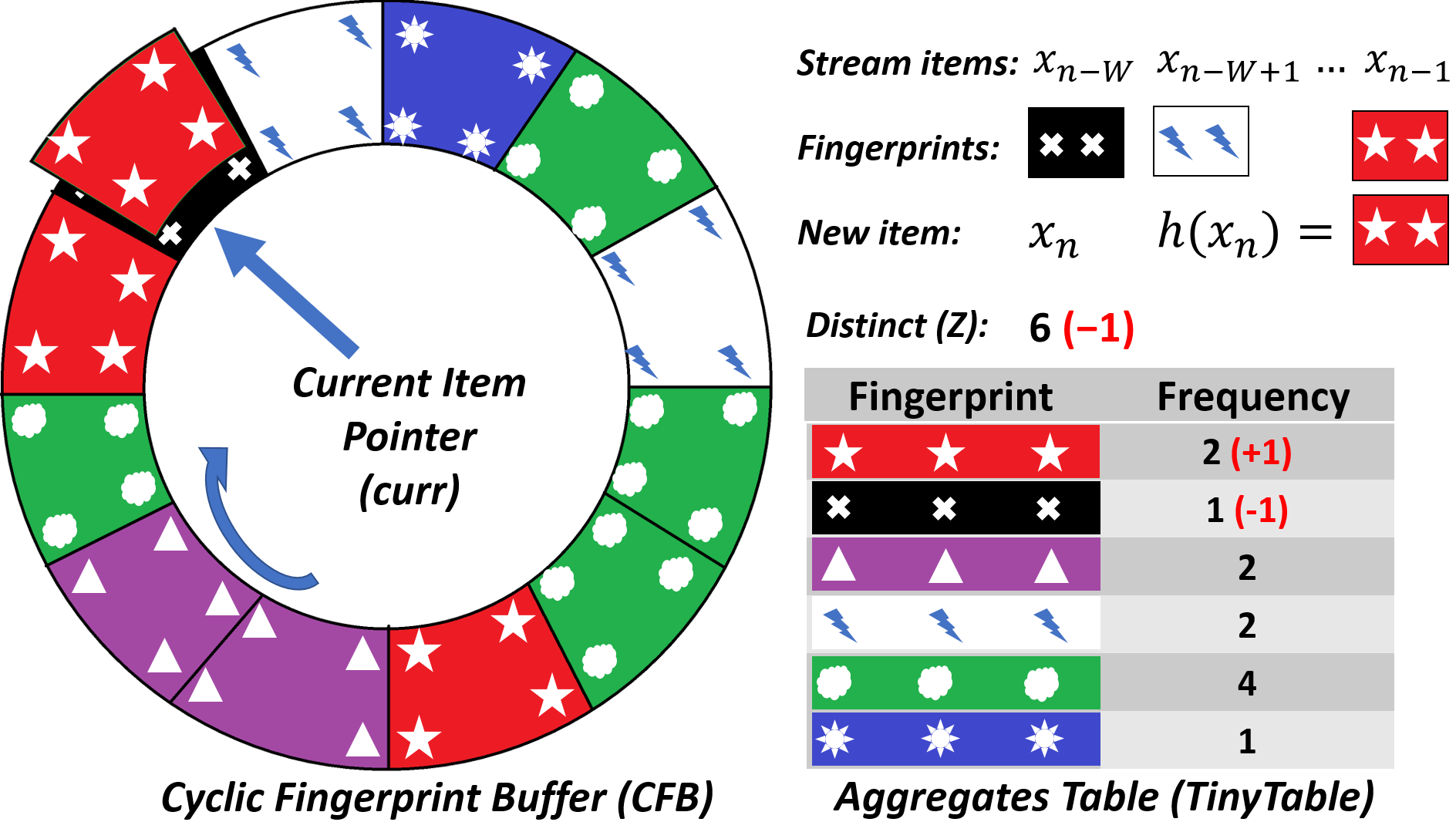}
		\caption{\label{fig:example} An overview of SWAMP: Fingerprints are stored in a cyclic fingerprint buffer (CFB), and their frequencies are maintained by TinyTable. Upon item $x_n$'s arrival, we update CFB and the table by removing the oldest item's ($x_{n-W}$) fingerprint (in black) and adding that of $x_n$ (in red).     We also maintain an estimate for the number of distinct fingerprints (Z). Since the black fingerprints count is now zero, we~decrement~Z.
	}}
\end{figure}

In this work, we suggest \emph{Sliding Window Approximate Measurement Protocol (SWAMP)}, an algorithm that bundles together four commonly used measurement types. Specifically, it approximates set membership, per flow counting, distinct elements and entropy in the sliding window model. As illustrated in Figure~\ref{fig:example}, SWAMP stores flows’ fingerprints\footnote{
A fingerprint is a short random string obtained by hashing an ID.} 
in a cyclic buffer while their frequencies are maintained in a compact fingerprint hash table named TinyTable~\cite{TinyTable}.
On each packet arrival, its corresponding fingerprint replaces the oldest one in the buffer.
We then update the table, decrementing the departing fingerprint's frequency and incrementing that of the arriving one.
An additional counter $Z$ maintains the number of distinct \textbf{fingerprints} in the window and is updated every time a fingerprint's frequency is reduced to $0$ or increased to $1$.
Intuitively, the number of distinct fingerprints provides a good estimation of the number of distinct elements.  Additionally, the scalar $\hat{H}$ (not illustrated) maintains the fingerprints’ distribution entropy and approximates the real entropy.


\subsection{Contribution}
We present \emph{SWAMP}, a sliding window algorithm for approximate set membership (Bloom filters), per-flow counting, distinct elements and entropy measurements.
We prove that SWAMP operates in constant time and provides accuracy guarantees for each of the supported problems.
Despite its versatility, SWAMP improves the state of the art for each.

For approximate set membership, SWAMP is memory succinct when the false positive rate is constant and requires up to 40\% less space than~\cite{slidngBloomFilterInfocom}.
SWAMP is also succinct for per-flow counting and is more accurate than~\cite{WCSS} on real packet traces.
When compared with $1+\varepsilon$ count distinct approximation algorithms~\cite{SlidingHLL,Fusy-HLL}, SWAMP asymptotically improves the query time from $O(\varepsilon^{-2})$ to a constant. It is also up to x1000 times more accurate on real packet traces.
For entropy, SWAMP asymptotically improves the runtime to a constant and provides accurate estimations in practice.

\begin{table}[h]
	\centering
	\scriptsize
	\begin{tabular}{|c|c|c|c|c|}
		\hline
		Algorithm  & Space  &  Time & Counts   \tabularnewline
		\hline
		\hline
		SWAMP  & $(1+o(1))\cdot W\log_2 W$  & $O\left(1\right)$  & \cmark  \tabularnewline
		\hline
		SWBF~\cite{slidngBloomFilterInfocom}  & $(2+o(1))\cdot W\log_2 W$  & $O\left(1\right)$  & \xmark \tabularnewline
		\hline
		TBF~\cite{TBF}  & $O\left(W\log_2 W\log_2\epsilon^{-1}\right)$  &$O\left(\log_2\epsilon^{-1}\right)$   & \xmark   \tabularnewline
		\hline
	\end{tabular}
	\begin{tabular}{|c|c|c|c|c|}
	\end{tabular}
	\normalsize
	\caption{
		Comparison of sliding window set membership
		\ifdefined\EXTENDED
		algorithms
		\fi
		for $\epsilon=W^{-o(1)}$.
	}
	\label{tbl:setMembership}
	\ifdefined\NINEPAGES
	\vspace*{-0.3cm}
	\fi
	
\end{table}

\normalfont
\begin{table*}[t!]
	\ifdefined\NINEPAGES
	\footnotesize
	\fi
	\centering{\hspace*{-0.5cm}
		\begin{tabular}{|c|c|c|c|}
			\hline
			Problem & Estimator
			& Guarantee
			& Reference
			\tabularnewline
			\hline
			\hline
			\multirow{2}{*}{\SetProblem} & \multirow{2}{*}{{\sc IsMember()}} & $\Pr(true|x\in S^W) =1$ & \multirow{2}{*}{Corollary~\ref{cor:bf}}\tabularnewline
			\cline{3-3}
			&  & $\Pr(true|x\notin S^W) \le \epsilon$ & \tabularnewline
			\hline
			\multirow{3}{*}{\SetMulProblem} & \multirow{2}{*}{{\sc Frequency} $(\widehat{f_x})$} & $\Pr \left(f_x \le \widehat{f_x}\right) =1$
			& \multirow{3}{*}{Theorem~\ref{thm:setmul}}\tabularnewline
			\cline{3-3}
			&  & $\Pr\left(f_x \neq \widehat{f_x} \right) \le \epsilon$ &\tabularnewline
			\hline
			\multirow{4}{*}{\CountDistinctProblem} & \multirow{2}{*}{\distinctLB{} $(Z)$} & $\Pr(D\ge Z)=1$ &\multirow{3}{*}{Theorem~\ref{thm:epsDeltaDistinctLB}}\tabularnewline
			\cline{3-3}
			&  & $\Pr \Big( { D - Z \ge \frac{1}{2}\varepsilon D \cdot \log \left( {\frac{2}{\delta }} \right)} \Big) \le \delta.$ &\tabularnewline
			\cline{2-4}
			& \distinctMLE{}$(\hat{D})$ & $\Pr \left( { \left|D - \hat{D}\right| \ge \frac{1}{2}\varepsilon D \cdot \log \left( {\frac{2}{\delta }} \right)} \right) \le \delta.$ &
			Theorem~\ref{thm:prob margin}
			
			\tabularnewline
			\hline
			\multirow{2}{*}{\EntropyProblem} & \multirow{2}{*}{{\sc Entropy} $(\hat{H})$} & $\Pr\left(H \ge \hat{H}\right) =1$ & \multirow{3}{*}{Theorem~\ref{thm:entropyInterval}}\tabularnewline \cline{3-3}
			&  &  $ \Pr \left ( H - \widehat{H} \ge \epsilon\delta^{-1} \right ) \le  \delta .$ &\tabularnewline
			\hline
		\end{tabular}
	}
	\caption{
		Summary of SWAMP's accuracy guarantees.
	}
	\ifdefined\NINEPAGES
	\vspace*{-0.7cm}
	\fi
	\label{tinytbl}
	\normalfont
\end{table*}

\ifdefined\EXTENDED
the notations we use can be found in Table~\ref{tbl:notations}.
\else

\fi

\ifdefined\NINEPAGES
\textbf{\textit{Roadmap.}}
\else
\subsection{Paper organization}
\fi
Related work on the problems covered by this work is found in Section~\ref{sec:related}.
Section~\ref{sec:SWAMP} provides formal definitions and introduces SWAMP.
Section~\ref{sec:Eval} describes an empirical evaluation of SWAMP and previously suggested algorithms.
Section~\ref{sec:anal} includes a formal analysis of SWAMP which is briefly summarized in Table~\ref{tinytbl}.
Finally, we conclude with a short discussion in Section~\ref{sec:discussion}.

\section{Related work}
\label{sec:related}
\subsection{Set Membership and Counting}
A Bloom filter~\cite{Bloom} is an efficient data structure that encodes an approximate set.
Given an item, a Bloom filter can be queried if that item is a part of the set.
An answer of `no' is always correct, while an answer of `yes' may be false with a certain probability.
This case is called \emph{False Positive}.

Plain Bloom filters do not support removals or counting and thus many algorithms fill this gap.
For example, some alternatives support removals~\cite{dleftCBF,TinyTable,TinySet,RankedIndexHashing,OceanStore,VLBF} and others support multiplicity queries~\cite{SpectralBloom,TinyTable}.
Additionally, some works use aging~\cite{ActiveActive} and others compute the approximate set with regard to a sliding windows~\cite{TBF,slidngBloomFilterInfocom}.

SWBF~\cite{slidngBloomFilterInfocom} uses a Cuckoo hash table to build a sliding Bloom filter, which is more space efficient than previously suggested \emph{Timing Bloom filters (TBF)~\cite{TBF}}.

The Cuckoo table is allocated with $2W$ entries such that each entry stores a fingerprint and a time stamp.
Cuckoo tables require that $W$ entries remain empty to avoid circles and this is done implicitly by treating cells containing outdated items as `empty'.
Finally, a cleanup process is used to remove outdated items and allow timestamps to be wrapped around.
A comparison of SWAMP, TBF and SWBF appears in Table~\ref{tbl:setMembership}.




\subsection{Count Distinct}
The number of \textbf{distinct} elements provides a useful indicator for anomaly detection algorithms.
Accurate count distinct is impractical due to the massive scale of the data~\cite{HLL} and thus most approaches resort to approximate solutions~\cite{CD1,CD2,CD3}.

Approximate algorithms typically use a hash function $H: \mathbb{ID}\to \{0,1\}^{\infty}$ that maps ids to infinite bit strings.
In practice, finite bit strings are used and $32$ bit integers suffice to reach estimations of over $10^9$~\cite{HLL}. These algorithms look for
certain \emph{observables} in the hashes.
For example, some algorithms~\cite{CD1,Giroire2009406} treat the minimal observed hash value as a real number in $[0,1]$ and exploit the fact that
$\mathbb{E}(\min\left(H(\cal{M})\right)) = \frac{1}{D+1}$, where $D$ is the real number of distinct items in the multi-set $\cal{M}$.
Alternatively, one can seek patterns of the form $0^{\beta-1}1$~\cite{CD3,HLL} and exploit the fact that such a pattern is encountered on average once per every $2^{\beta}$ unique elements.

Monitoring observables reduces the required amount of space as we only need to maintain a single one.
In practice, the variance of such methods is large and hence multiple observables are maintained.

In principle, one could repeat the process and perform $m$ independent experiments but this has significant computational overheads.
Instead, \emph{stochastic averaging}~\cite{CD4} is used to mimic the effects of multiple experiments with a single hash calculation. At any case, using  $m$ repetitions reduces the standard deviation by a factor of~$\frac{1}{\sqrt{m}}$.

The state of the art count distinct algorithm is~\emph{HyperLogLog (HLL)}~\cite{HLL}, which is used in multiple Google projects~\cite{HLLInPractice}.
HLL requires $m$ bytes and its standard deviation is $\sigma \approx \frac{1.04}{\sqrt{m}}$.
SWHLL extends HLL to sliding windows~\cite{SlidingHLL,Fusy-HLL}, and was used to detect attacks such as port scans~\cite{Chabchoub2014}.
SWAMP's space requirement is proportional to $W$ and thus, it is only comparable in space to HLL when $\varepsilon^{-2} = O(W)$.
However, when multiple functionalities are required the \emph{residual} space overhead of SWAMP is only $\log(W)$ bits, which is considerably less than any standalone alternative.
\subsection{Entropy Detection}
Entropy is commonly used as a signal for anomaly detection~\cite{Entropy1}.
Intuitively, it can be viewed as a summary of the entire traffic histogram.
The benefit of entropy based approaches is that they require no exact understanding of the attack's mechanism.
Instead, such a solution assumes that a sharp change in the entropy is caused by~anomalies.


An $\epsilon,\delta$ approximation of the entropy of a stream can be calculated in $O\left( {{\varepsilon ^{ - 2}}\log {\delta ^{ - 1}}} \right)$ space~\cite{SudiptoAndMcGregor}, an algorithm that was also extended to sliding window using priority sampling~\cite{PrioritySampling}.
That sliding window algorithm is improved by~\cite{OptimalSamplingSW} whose algorithm requires $O\left( {{\varepsilon ^{ - 2}}\log {\delta ^{ - 1}}\log \left( N \right)} \right)$ memory.

\subsection{Preliminaries -- Compact Set Multiplicity}
\label{sec:TinyTable}
Our work requires a compact set multiplicity structure that support both set membership and multiplicity queries. TinyTable~\cite{TinyTable} and CQF~\cite{CQF}  
fit the description while other structures~\cite{dleftCBF,RankedIndexHashing} are naturally expendable for multiplicity queries with at the expense of additional space. We choose TinyTable~\cite{TinyTable} as its code is publicly available as open source.

TinyTable encodes $W$ fingerprints of size $L$ using $\left( {1 + \alpha } \right)W\left( {L - \log_2 \left( W \right) + 3} \right) + o\left( W \right)$ bits, where $\alpha$ is a small constant that affects update speed; when $\alpha$ grows, TinyTable becomes faster but also consumes more space.

\section{SWAMP Algorithm}
\label{sec:SWAMP}

\subsection{Model}
We consider a stream  $(\mathbb{S})$ of IDs where at each step an ID is added to $\mathbb{S}$.
The last $W$ elements in $\mathbb{S}$ are denoted $\mathbb{S}^{W}$.
Given an ID $y$, the notation $f_y^{W}$ represents the frequency of $y$ in $\mathbb{S}^{W}$.
Similarly, $\widehat{f_y^{W}}$ is an approximation of $f_y^{W}$.
For ease of reference, notations are summarized in Table~\ref{tbl:notations}.

\begin{table}[h]
	\centering{
	\begin{tabular}{|c|l|}
		
		\hline
		Symbol & Meaning \tabularnewline
		\hline
		$W$ & Sliding window size \tabularnewline
		\hline
		$\mathbb{S}$ & Stream. \tabularnewline
		\hline
		$\mathbb{S}^{W}$ & Last $W$ elements in the stream. \tabularnewline
		\hline
		$\varepsilon$ & Accuracy parameter for sets membership.
		\tabularnewline
		\hline
		$\varepsilon_D$ & Accuracy parameter for count distinct.
		\tabularnewline
			\hline
		$\varepsilon_H$ & Accuracy parameter for entropy estimation.
		\tabularnewline
		\hline
		$\delta$ & Confidence for count distinct.
		\tabularnewline
		\hline
		$L$ & Fingerprint length in bits, $L \triangleq\log_2\left(W\varepsilon^{-1}\right)$.  \tabularnewline
		\hline
		$f_y^{W}$ & Frequency of ID $y$ in $\mathbb{S}^{W}$.
		\tabularnewline
		\hline
		$\widehat{f_y^{W}}$ & An approximation of $f_y^{W}$.
		\tabularnewline
		\hline
		$D$ & Number of distinct elements in $\mathbb{S}^{W}$.
		\tabularnewline
		\hline	
		$Z$ & Number of distinct fingerprints.
		\tabularnewline
		\hline	
		$\hat{D}$ & Estimation provided by DISTINCTMLE function.
		\tabularnewline
		\hline			
		$\alpha$ & TinyTable's parameter ($\alpha = 1.2$).
		\tabularnewline
		\hline		
		$h$ & A pairwise independent hash function.
		\tabularnewline
		\hline	
			$F$ & A set of fingerprints stored in $CFB$. 
		\tabularnewline
		\hline				
	\end{tabular}
}
	\caption{List of symbols}
	\label{tbl:notations}
\ifdefined\NINEPAGES	
\vspace*{-0.7cm}
\fi	
\end{table}
\normalsize 
\begin{algorithm}[t!]
	\caption{SWAMP}
	\footnotesize
	\ifdefined\NINEPAGES
	\footnotesize
	\fi
	\begin{algorithmic}[1]
		\ifdefined\EXTENDED
		\Require{TinyTable $TinyTable$, Fingerprint Array $CFB$, integer $curr$, integer $Z$ }
		
		\Statex \textbf{initialization}
		\Statex $CFB \gets \bar 0$ 
		\Statex $curr \gets 0$  \Comment{Initial index is 0}.
		\Statex $Z \gets 0$  \Comment{0 distinct fingerprints}.
		\Statex $TT\gets TinyTable$
		\Statex $\entropyVariable\gets 0$ \Comment{0 entropy}
		\else
		\Statex \textbf{init: $CFB \gets \bar 0, curr \gets 0, Z \gets 0, TT\gets TinyTable, \entropyVariable\gets 0$}
		\fi
		\Function{Update }{ID $x$}
		\State $prevFreq \gets TT.frequency(CFB[curr])$
		\State $TT.remove(CFB[curr])$
		\State $CFB[curr] \gets h(x)$
		\State $TT.add(CFB[curr])$
		\State $xFreq \gets TT.frequency(CFB[curr])$
		\State \Call {UpdateCD}{$prevFreq,xFreq$}
		\State \Call {UpdateEntropy}{$prevFreq,xFreq$}
		\State $curr \gets (curr +1)$ \textbf{mod} $W$
		
		\EndFunction
		\Procedure{UpdateCD}{$prevFreq,xFreq$}
		\If {$prevFreq = 1$}
		\State $Z \gets Z -1$
		\EndIf
		\If {$xFreq = 1$}
		\State $Z \gets Z +1$
		\EndIf
		\EndProcedure
		\Procedure{UpdateEntropy}{$prevFreq,xFreq$}
		\State $PP \gets \frac{prevFreq}{W}$ \Comment{The previous probability}
		\State $CP \gets \frac{prevFreq-1}{W}$ \Comment{The current probability}
		\State $\entropyVariable \gets \entropyVariable +PP \logp{PP} - CP\logp{CP}$
		\State $xPP \gets \frac{xFreq-1}{W}$ \Comment{$x$'s previous probability}
		\State $xCP \gets \frac{xFreq}{W}$ \Comment{$x$'s current probability}
		\State $\entropyVariable \gets \entropyVariable + xPP \logp{xPP} - xCP\logp{xCP}$
		\EndProcedure
		\Function {IsMember}{ID $x$}
		\If {$TT.frequency(h(x))>0$}
		\State \Return true \EndIf
		\State \Return false
		\EndFunction
		\Function{Frequency }{ID $x$}
		\State \Return $TT.frequency(h(x))$
		\EndFunction
		\Function {\distinctLB()}{}
		\State \Return $Z$
		\EndFunction
		\Function {\distinctMLE()}{}
		\State $\hat{D} \gets \frac{\ln\left(1 - \frac{Z}{2^L}\right)}{\ln \left(1 - \frac{1}{2^L} \right)}$
		\State \Return $\hat{D}$
		\EndFunction
		\Function {\entropy()}{}
		\State \Return $\entropyVariable$
		\EndFunction
	\end{algorithmic}
	\label{alg:SWAMP}
	\normalsize
\end{algorithm}

\subsection{Problems definitions}
We start by formally defining the approximate set membership problem.
\begin{definition}
	We say that an algorithm solves the \SetProblem{} problem if given an ID $y$, it returns true if $y\in \mathbb{S}^{W}$ and if $y \notin \mathbb{S}^{W}$, it returns false with probability of at least $1-\varepsilon$.
\end{definition}
The above problem is solved by SWBF~\cite{slidngBloomFilterInfocom} and by \emph{Timing Bloom filter (TBF)}~\cite{TBF}.
In practice, SWAMP solves the stronger \SetMulProblem{} problem, as defined below:

\begin{definition}
	We say that an algorithm solves the \SetMulProblem{} problem if given an ID $y$, it returns an estimation $\widehat{f_y^{W}}$ s.t. $\widehat{f_y^{W}}\ge f_y^{W}$ and with probability of at least $1 -\varepsilon$:
	${f_y}^W = \widehat{{f_y}^W}$.
\end{definition}

Intuitively, the \SetMulProblem{} problem guarantees that we always get an over approximation of the frequency and that with probability of at least $1 -\varepsilon$ we get the exact window frequency.
A simple observation shows that any algorithm that solves the \SetMulProblem{} problem also solves the \SetProblem{} problem.
Specifically, if $y \in \mathbb{S}^{W}$, then $\widehat{f_y^{W}}\ge f_y^{W}$ implies that $\widehat{f_y^{W}}\ge 1$ and we can return true.
On the other hand, if $y \notin \mathbb{S}^{W}$, then ${f_y}^W = 0$ and with probability of at least $1-\varepsilon$, we get: ${f_y}^W = 0 = \widehat{{f_y}^W}$.
Thus, the \emph{isMember} estimator simply returns true if $\widehat{{f_y}^W}>0$ and false otherwise. We later show that this estimator solves the \SetProblem{} problem.

\begin{figure*}[t!]
	\center{
		\subfigure[\label{subfig:SBF}Sliding Bloom filter ]{\includegraphics[width = 5.6cm]{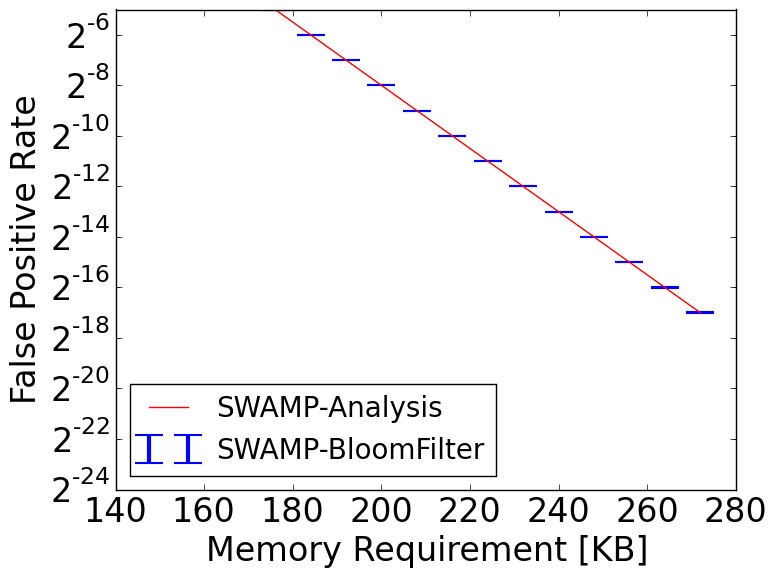}}
		\subfigure[\label{subfig:Z}Count Distinct]{\includegraphics[width = 5.6cm]{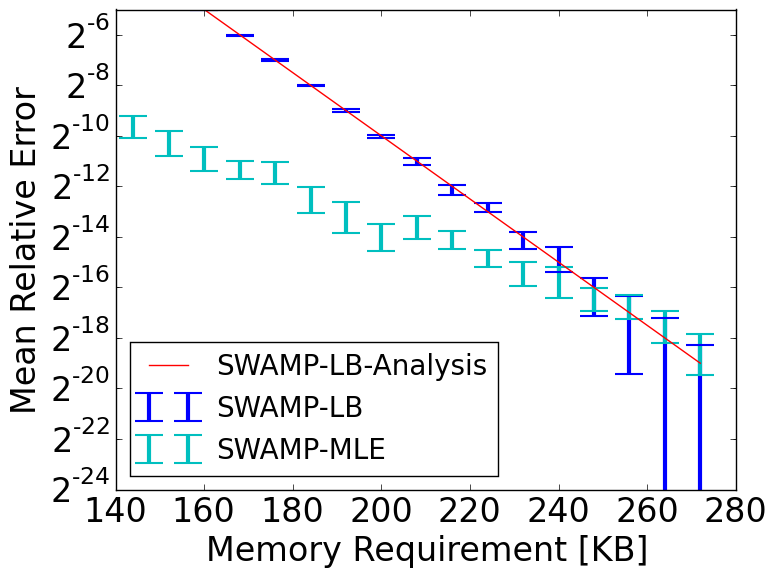}}
		\subfigure[\label{subfig:entropy}Entropy]{\includegraphics[width = 5.6cm]{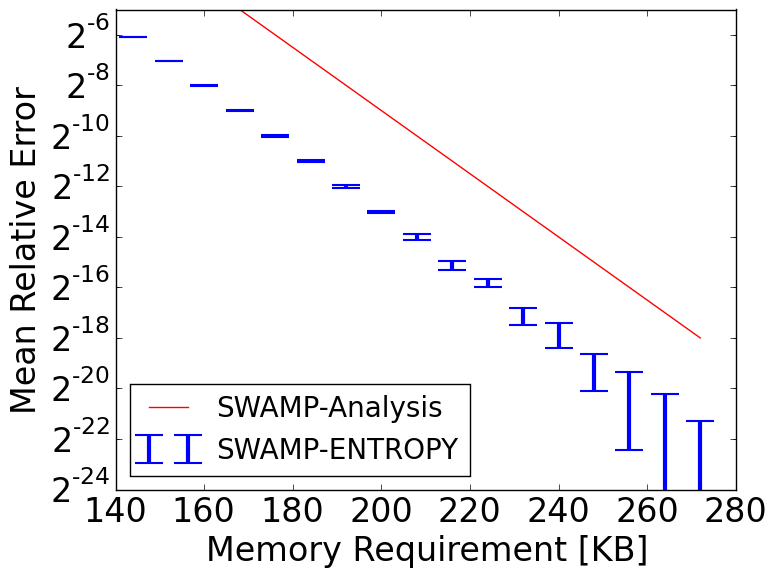}}
	}
	\caption{Empirical error and theoretical guarantee for multiple functionalities (random inputs).}
	\label{fig:anal}
\end{figure*}
The goal of the \CountDistinctProblem{} problem is to maintain an estimation of the number of distinct elements in $\mathbb{S}^{W}$.
We denote their number by $D$.
\begin{definition}
	We say that an algorithm solves the \CountDistinctProblem{} problem if it returns an estimation $\widehat{D}$ such that
	$D \ge \widehat{D}$ and with probability $1-\delta$: $\widehat{D} \ge \left( {1 - {\varepsilon}} \right)D $.
\end{definition}

Intuitively, an algorithm that solves the \CountDistinctProblem{} problem is able to conservatively estimate the number of distinct elements in the window and with probability of $1-\delta$, this estimate is close to the real number of distinct elements.

The entropy of a window is defined as: \[H \triangleq  - \sum\nolimits_{i = 1}^D {\frac{{{f_i}}}{W}} \log \left( {\frac{{{f_i}}}{W}} \right),\]
where $D$ is the number of distinct elements in the window, $W$ is the total number of packets, and $f_i$ is the frequency of flow $i$. We define the window entropy estimation problem as:
\begin{definition}
	An algorithm solves the \EntropyProblem{} problem, if it provides an estimator $\hat{H}$ so that $H\ge\hat{H}$ and $\Pr\left(H - \hat{H} \ge \epsilon\right)\le \delta$.
\end{definition}

\subsection{SWAMP algorithm}

We now present~\emph{Sliding Window Approximate Measurement Protocol (SWAMP)}.
SWAMP uses a single hash function ($h$), which given an ID $(y)$, generates $L\triangleq \left\lceil\log_2(W\epsilon^{-1})\right\rceil$ random bits $h(y)$ that are called its \emph{fingerprint}. We note that $h$ only needs to be pairwise-independent and can thus be efficiently implemented using only $O(\log W)$ space.
Fingerprints are then stored in a cyclic fingerprint buffer of length $W$ that is denoted $CFB$. The variable $curr$ always points to the oldest entry in the buffer.
Fingerprints are also stored in TinyTable~\cite{TinyTable} that provides compact encoding and multiplicity information.

\begin{figure*}[t]
	\center{
		\ifdefined\EXTENDED
		\subfigure[\label{variableEps}Window size is $2^{16}$ and varying $\varepsilon$.]{\includegraphics[width =.49\columnwidth]{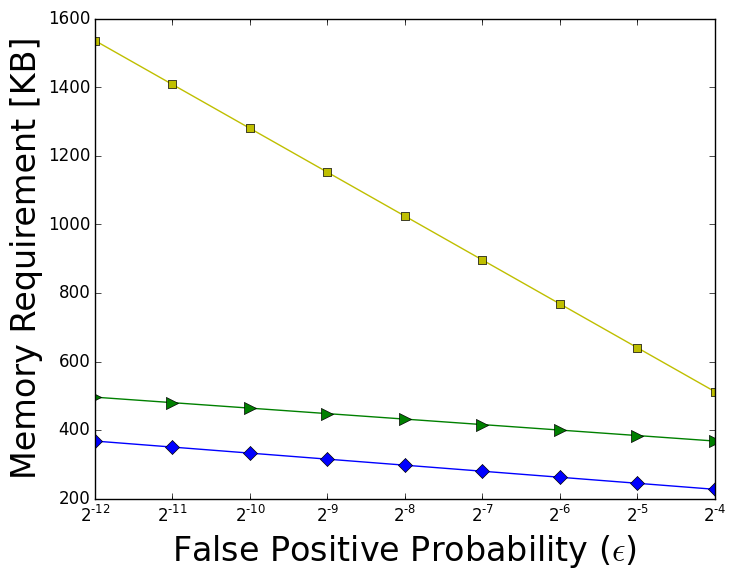}}
		\subfigure[\label{variableW}$\varepsilon =2^{-10}$ and varying window sizes.]{\includegraphics[width =.49\columnwidth]{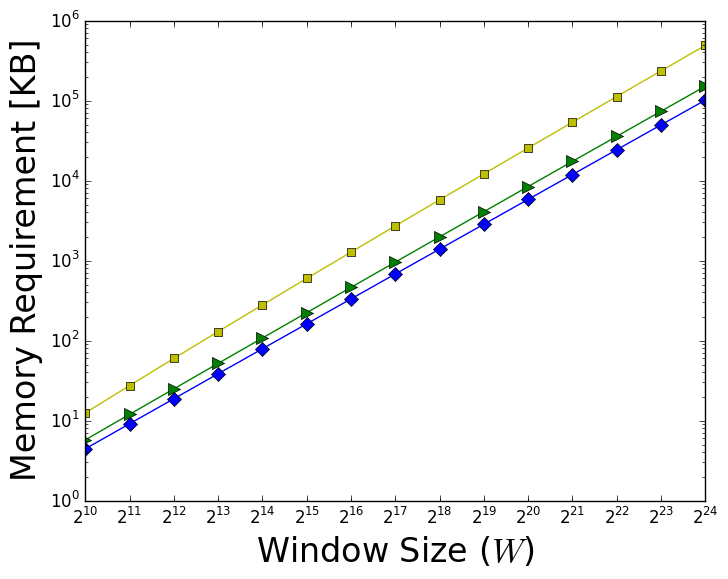}}
		\subfigure{\includegraphics[width = 7.8cm]{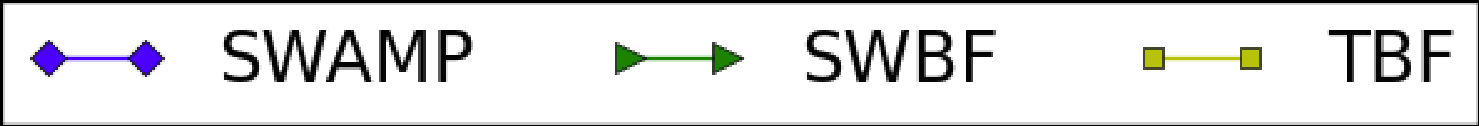}}
		\else
\ifdefined\EXTENDED		
		\subfigure[\label{variableEps}Window size is $2^{16}$ and varying $\varepsilon$.]{\includegraphics[width =\columnwidth]{W=65536_SlidingBloomFilter_EpsAxis.png}}
\else
		\subfigure[\label{variableEps}Window size is $2^{16}$ and varying $\varepsilon$.]{\includegraphics[width =.96\columnwidth]{W=65536_SlidingBloomFilter_EpsAxis.png}}
\fi		
\ifdefined\EXTENDED
		\subfigure[\label{variableW}$\varepsilon =2^{-10}$ and varying window sizes.]{\includegraphics[width =\columnwidth]{eps=00009765625_SlidingBloomFilter_WinSizeAxis.png}}
		\subfigure{\includegraphics[width = 7.8cm]{BFLegend.PNG}}
\else
		\subfigure[\label{variableW}$\varepsilon =2^{-10}$ and varying window sizes.]{\includegraphics[width =.96\columnwidth]{eps=00009765625_SlidingBloomFilter_WinSizeAxis.png}}
		\subfigure{\includegraphics[width = 7.8cm]{BFLegend.PNG}}
\fi		
		\fi
	}
	
	\caption{Memory consumption of sliding Bloom filters as a function of $W$ and $\varepsilon.$}
	\label{fig:W}
\end{figure*}
The update operation replaces the oldest fingerprint in the window with that of the newly arriving item.
To do so, it updates both the cyclic fingerprint buffer ($CFB$) and TinyTable.
In CFB, the fingerprint at location $curr$ is replaced with the newly arriving fingerprint.
In TinyTable, we remove one occurrence of the oldest fingerprint and add the newly arriving fingerprint.
The update method also updates the variable $Z$, which measures the number of distinct fingerprints in the window.
$Z$ is incremented every time that a new unique fingerprint is added to TinyTable, i.e., $FREQUENCY(y)$ changes from $0$ to $1$, where the method $FREQUENCY(y)$ receives a fingerprint and returns its frequency as provided by TinyTable.
Similarly, denote $x$ the item whose fingerprint is removed; if $FREQUENCY(x)$ changes to $0$, we decrement~$Z$.

SWAMP has two methods to estimate the number of distinct flows.
\distinctLB{} simply returns $Z$, which yields a conservative estimator of $D$ while \distinctMLE{} is an approximation of its Maximum Likelihood Estimator.
Clearly, \distinctMLE{} is more accurate than \distinctLB{}, but its estimation error is two sided.
A pseudo code is provided in Algorithm~\ref{alg:SWAMP} and an illustration is given by Figure~\ref{fig:example}.

\begin{figure*}[t!]
	\center{
		\subfigure[Chicago 16 Dataset]{\includegraphics[width = 5.6cm]{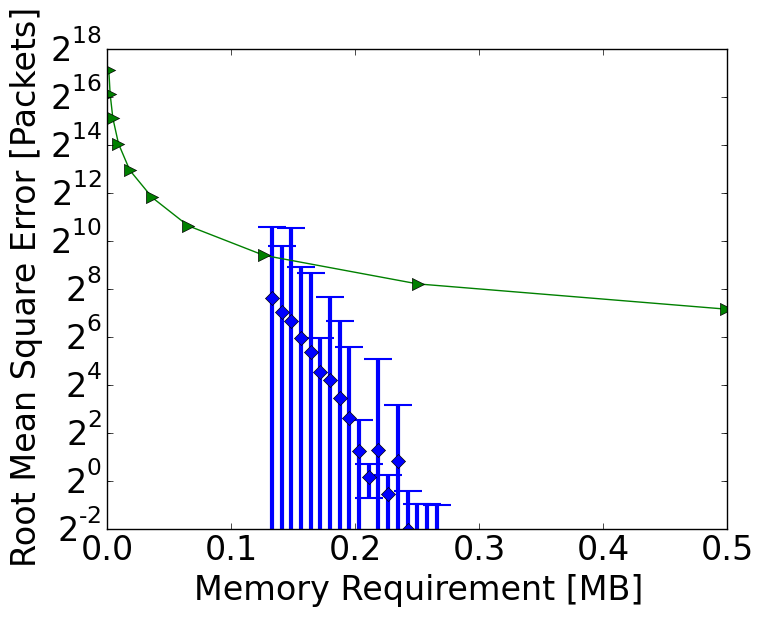}}
		\subfigure[San Jose 14 Dataset]{\includegraphics[width = 5.6cm]{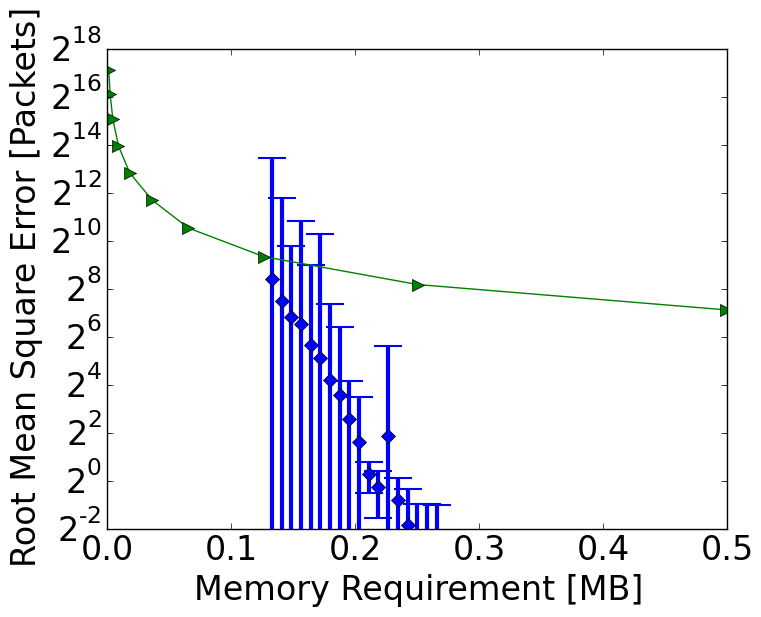}}
		\subfigure[DC 1 Dataset]{\includegraphics[width = 5.6cm]{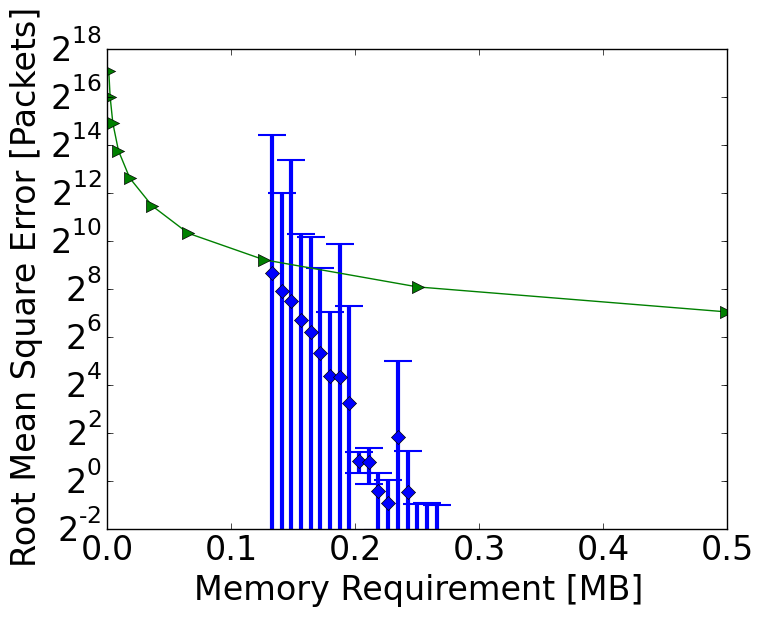}}		
		\subfigure{\includegraphics[width = 5.6cm]{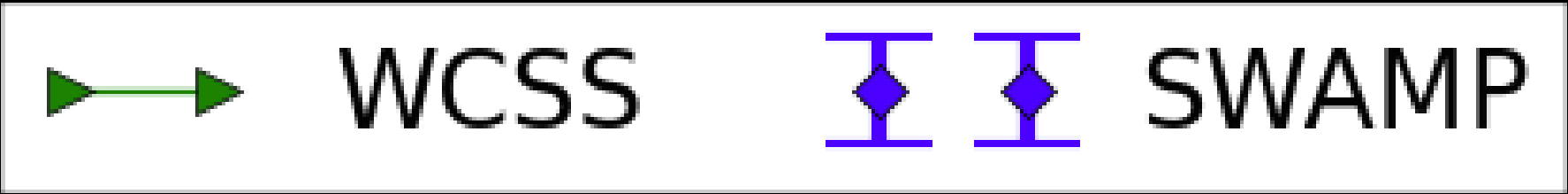}}\\
\ifdefined\EXTENDED		
		\addtocounter{subfigure}{-1}
		\subfigure[Chicago 15 Dataset]{\includegraphics[width = 5.6cm]{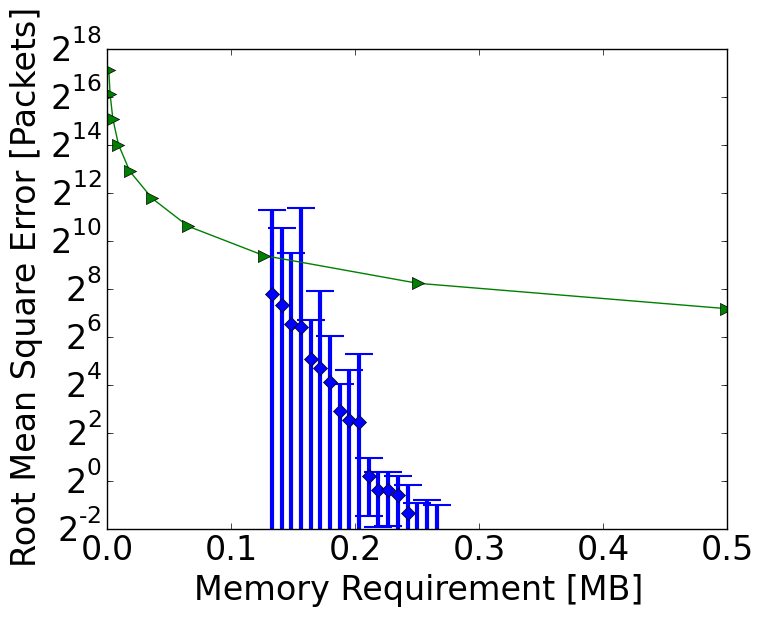}}
		\subfigure[San Jose 13 Dataset]{\includegraphics[width = 5.6cm]{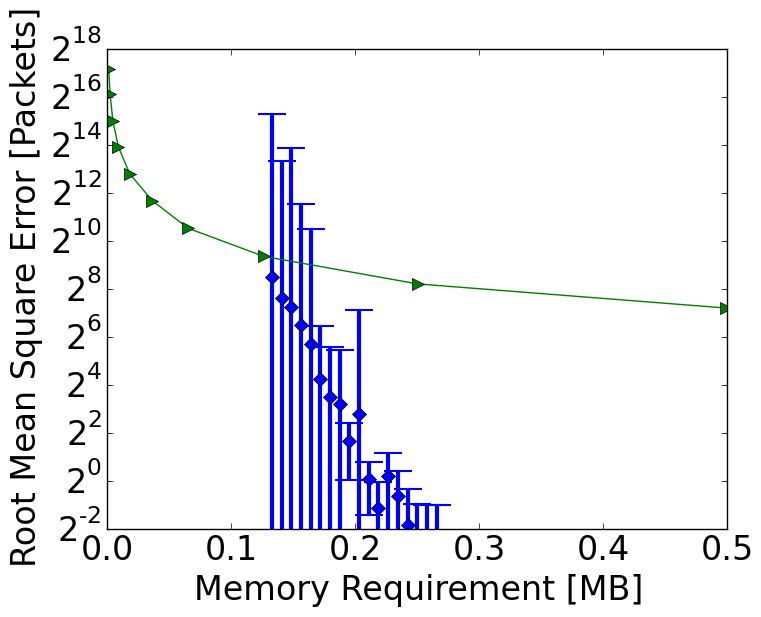}}
		\subfigure[DC 2 Dataset]{\includegraphics[width = 5.6cm]{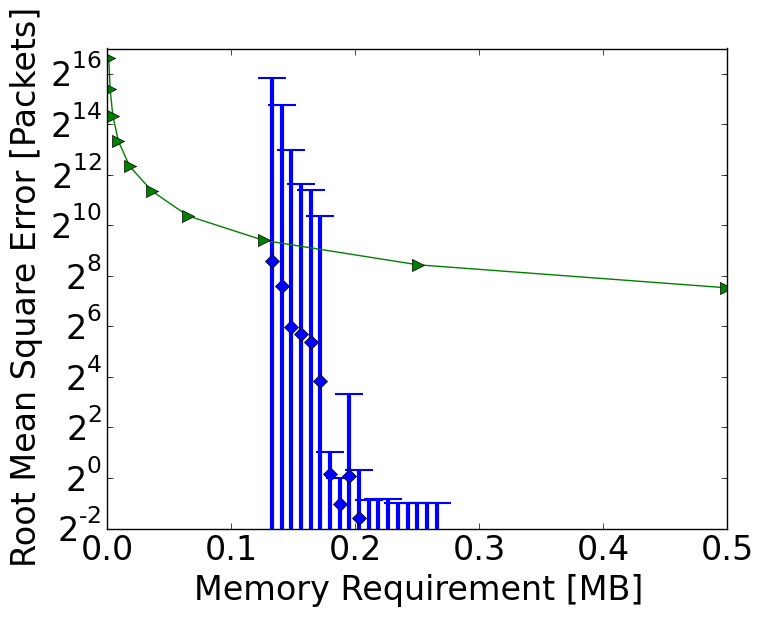}}
\fi	
	}
	\caption{Space/accuracy trade-off in estimating per flow packet counts (window size:  $W = 2^{16}$). }
	\ifdefined\NINEPAGES	
	\vspace*{-0.3cm}
	\fi
	\label{fig:memWCSS}
\end{figure*}
\begin{figure*}[t]
	\center{
		\subfigure[Chicago 16 Dataset]{\includegraphics[width = 5.6cm]{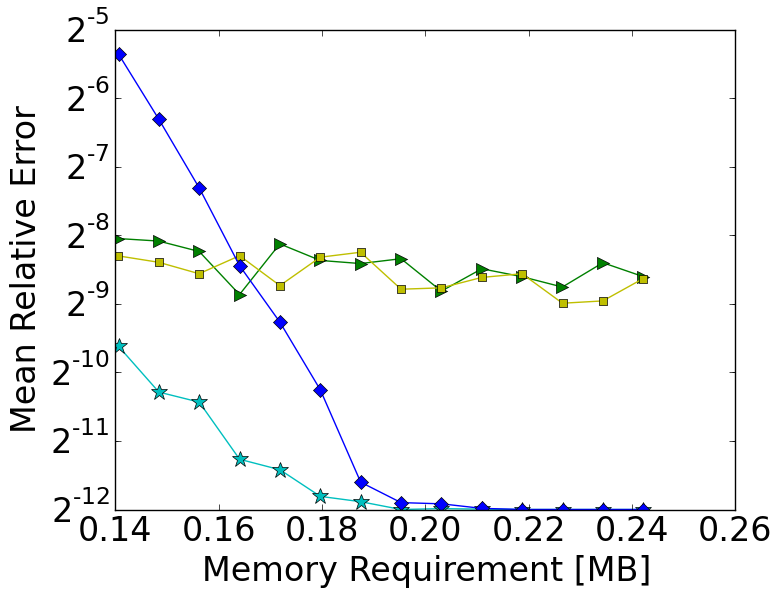}}
		\subfigure[San Jose 14 Dataset]{\includegraphics[width = 5.6cm]{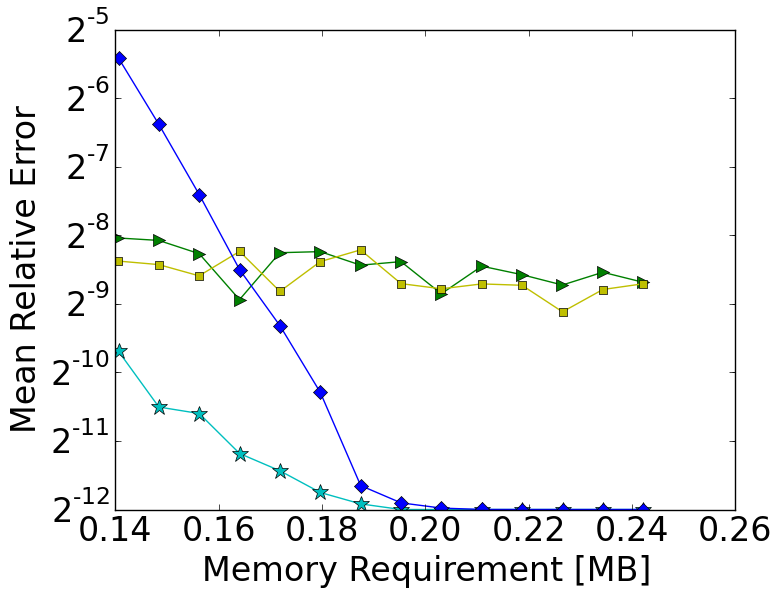}}
		\subfigure[DC 1 Dataset]{\includegraphics[width = 5.6cm]{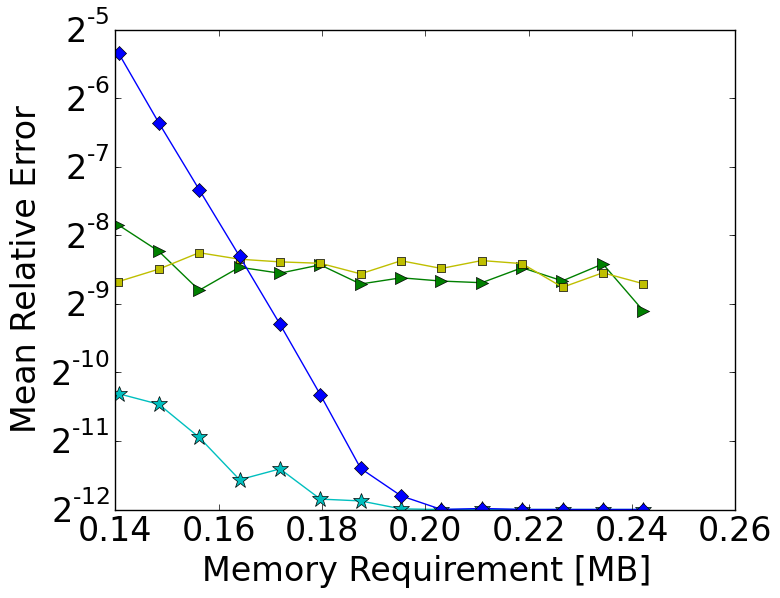}}		
		\subfigure{\includegraphics[width = 13.8cm]{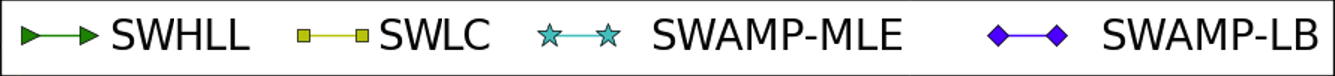}}
\ifdefined\EXTENDED		
		\addtocounter{subfigure}{-1}
		\subfigure[Chicago 15 Dataset]{\includegraphics[width = 5.6cm]{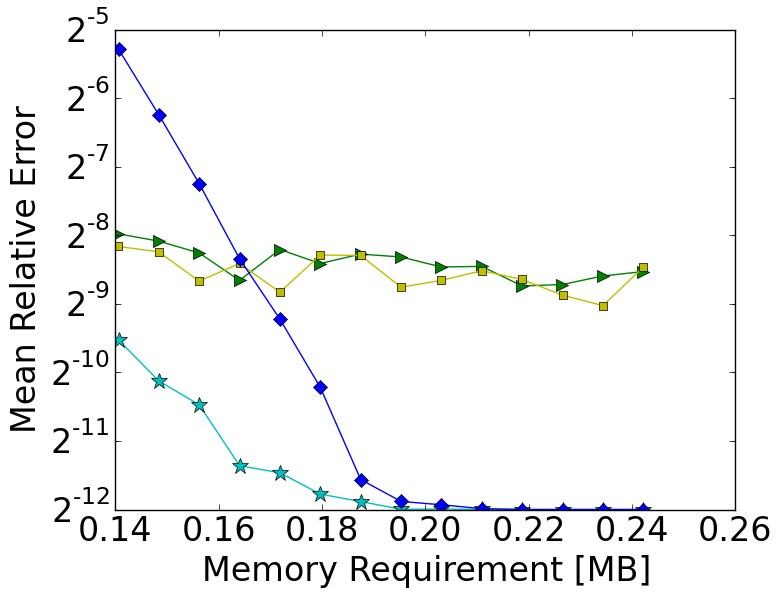}}
		\subfigure[San Jose 13 Dataset]{\includegraphics[width = 5.6cm]{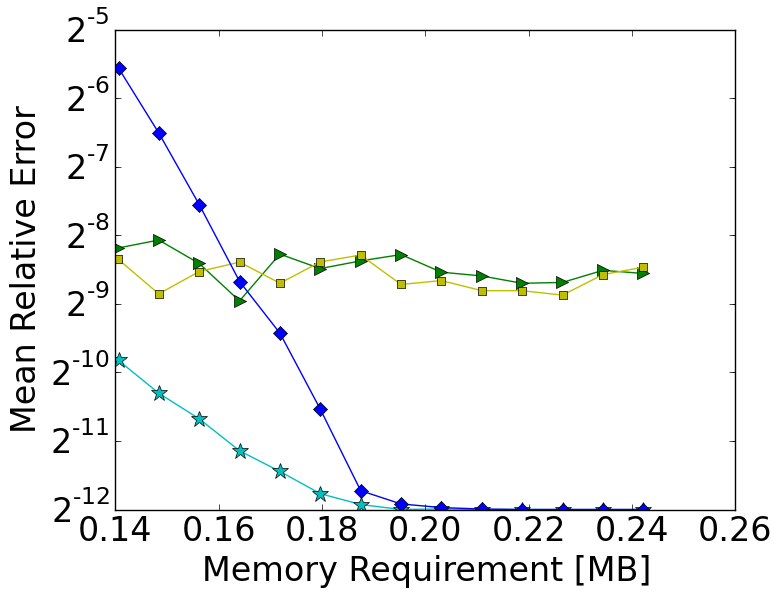}}
		\subfigure[DC 2 Dataset]{\includegraphics[width = 5.6cm]{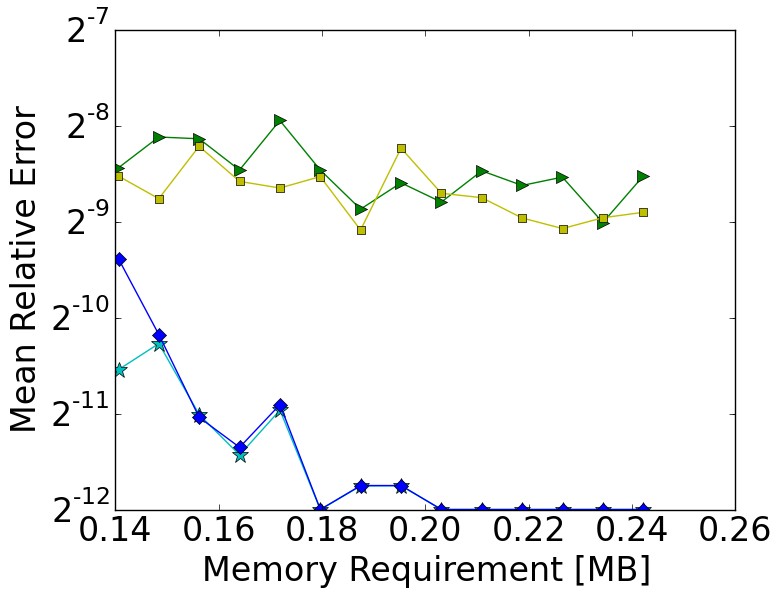}}
\fi		
	}
	\caption{Accuracy of SWAMP's count distinct functionality compared to alternatives.}
	\label{fig:Z}
\end{figure*}
\begin{figure*}[t]
	\center{
		\subfigure[Chicago 16 Dataset]{\includegraphics[width = 5.6cm]{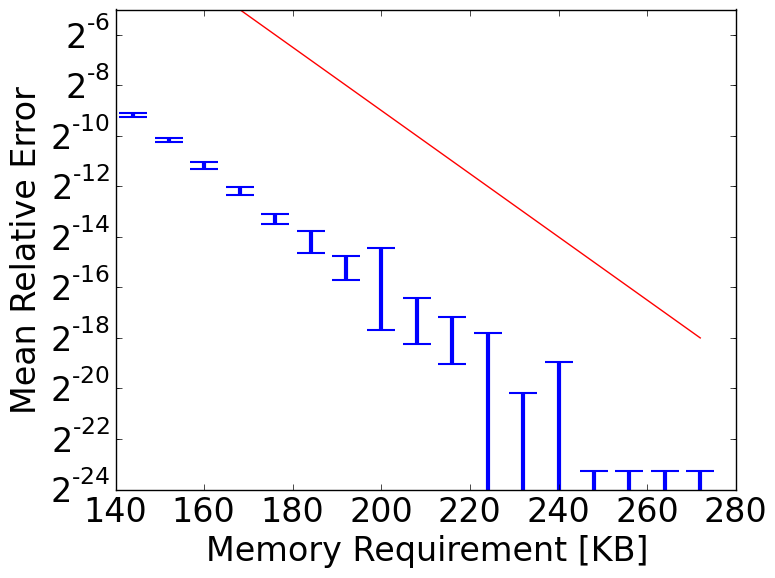}}
		\subfigure[San Jose 14 Dataset]{\includegraphics[width = 5.6cm]{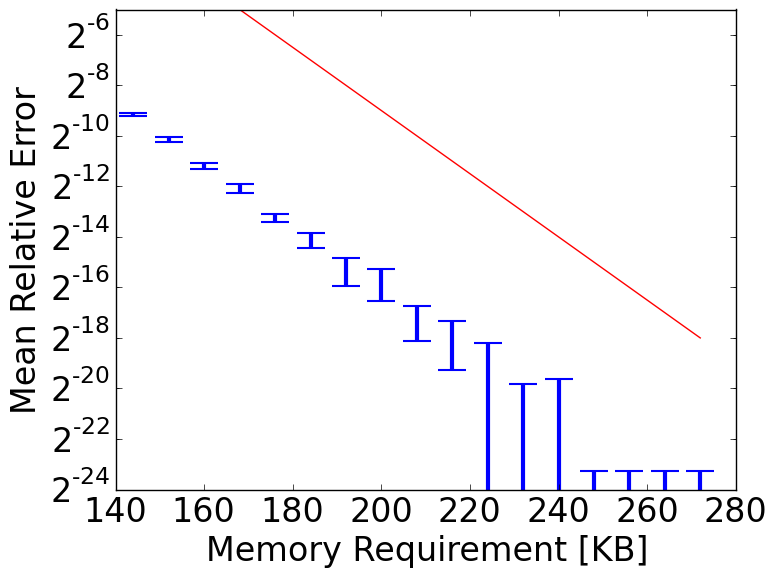}}
		\subfigure[DC 1 Dataset]{\includegraphics[width = 5.6cm]{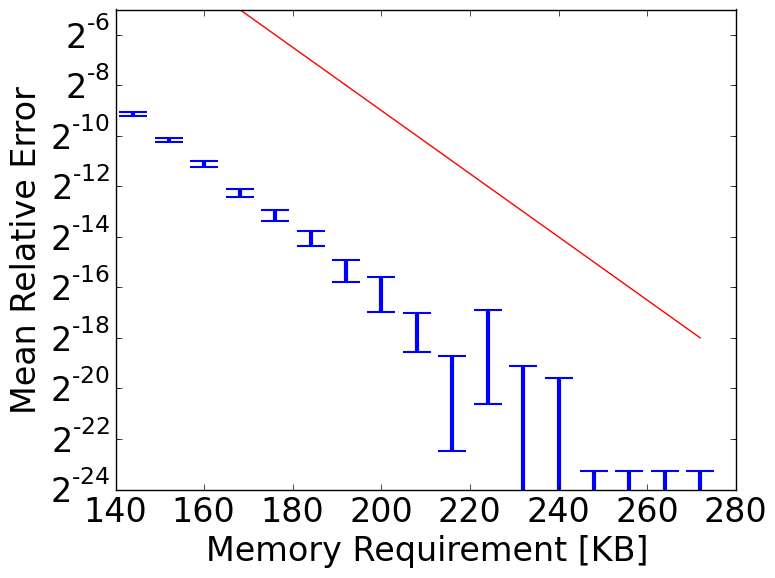}}		
		\subfigure{\includegraphics[width = 10.8cm]{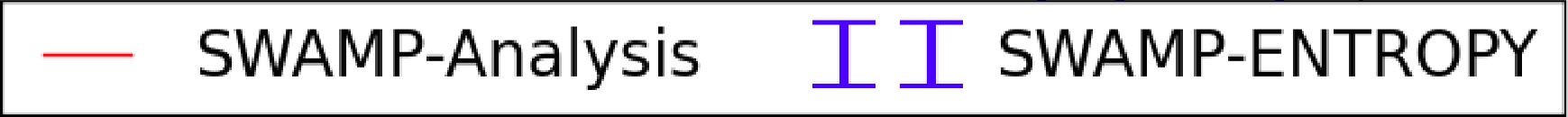}}\\
\ifdefined\EXTENDED
		\addtocounter{subfigure}{-1}
		\subfigure[Chicago 15 Dataset]{\includegraphics[width = 5.6cm]{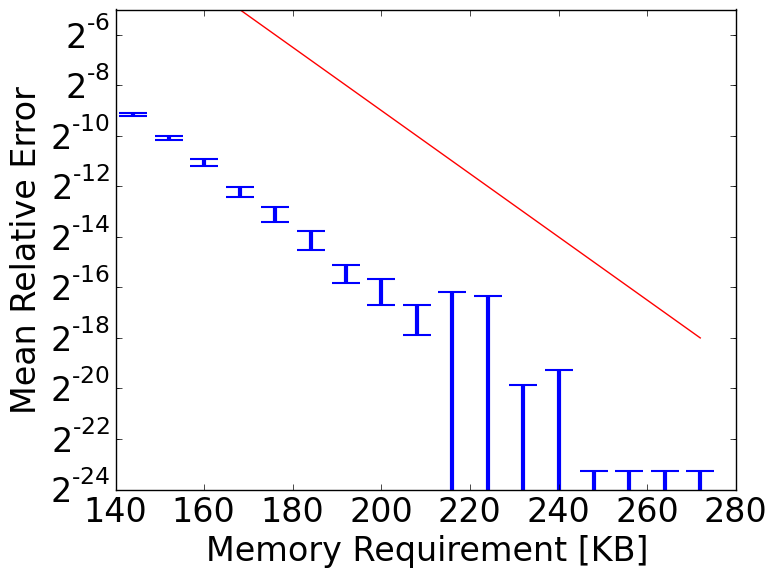}}
		\subfigure[San Jose 13 Dataset]{\includegraphics[width = 5.6cm]{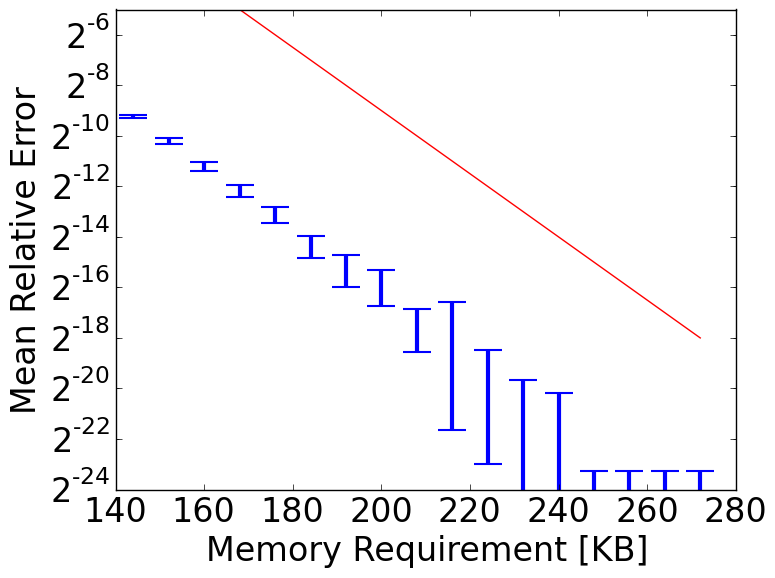}}
		\subfigure[DC 2 Dataset]{\includegraphics[width = 5.6cm]{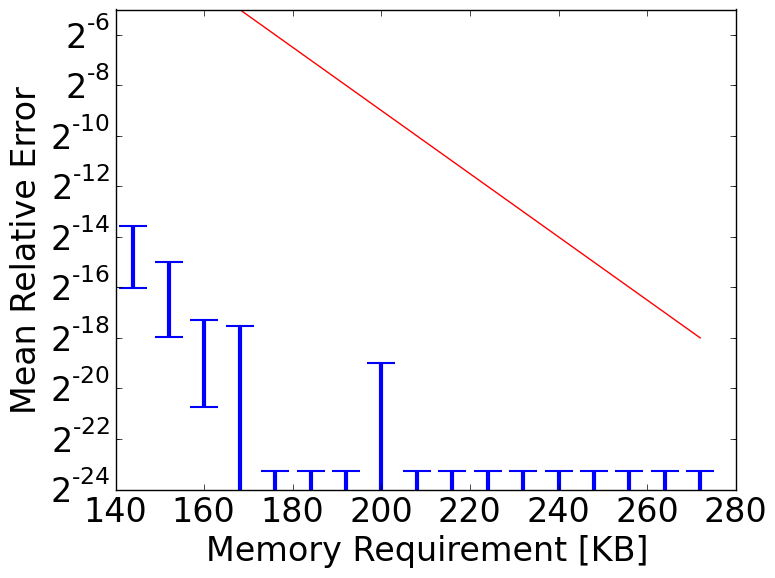}}
\fi		
	}
	\caption{Accuracy of SWAMP's entropy functionality compared to the theoretical guarantee.}
\ifdefined\NINEPAGES	
	\vspace*{-0.3cm}
\fi
	\label{fig:Entropy}
\end{figure*}
\section{Empirical Evaluation}
\label{sec:Eval}
\subsection{Overview}
We evaluate SWAMP's various functionalities, each against its known solutions.
We start with the \SetProblem{} problem where we compare SWAMP to SWBF~\cite{slidngBloomFilterInfocom} and \emph{Timing Bloom filter (TBF)}~\cite{TBF}, which only solve \SetProblem.

For counting, we compare SWAMP to~\emph{Window Compact Space Saving (WCSS)}~\cite{WCSS} that solves heavy hitters identification on a sliding window.
WCSS provides a different accuracy guarantee and therefore we evaluate their empirical accuracy when both algorithms are given the same amount of space.

For the distinct elements problem, we compare SWAMP against \emph{Sliding Hyper Log Log}~\cite{SlidingHLL,Fusy-HLL}, denoted SWHLL, who proposed running HyperLogLog and LogLog on a sliding window.
In our settings, small range correction is active and thus HyperLogLog and LogLog devolve into the same algorithm.
Additionally, since we already know that small range correction is active, we can allocate only a single bit (rather than 5) to each counter.
This option, denoted SWLC, slightly improves the space/accuracy ratio.

In all measurements, we use a window size of $W=2^{16}$ unless specified otherwise.
In addition, the underlying TinyTable uses $\alpha =0.2$ as recommended by its authors~\cite{TinyTable}.

Our evaluation includes six Internet packet traces consisting of backbone routers and data center traffic.
The backbone traces contain a mix of 1 billion UDP, TCP and ICMP packets collected from two major routers in Chicago~\cite{CAIDACH16} and San Jose~\cite{CAIDASJ14} during the years 2013-2016.
The dataset Chicago 16 refers to data collected from the Chicago router in 2016, San Jose 14 to data collected from the San Jose router in 2014, etc.
The datacenter packet traces are taken from two university datacenters that consist of up to 1,000 servers~\cite{Benson}. These traces are denoted DC1 and DC2.
\ifdefined\NINEPAGES
Due to lack of space, results for DC2 as well as additional Caida traces are deferred to the full version of this paper~\cite{full-version}.
\fi
\subsection{Evaluation of analytical guarantees}
Figure~\ref{fig:anal} evaluates the accuracy of our analysis from Section~\ref{sec:anal} on random inputs. 
As can be observed, the analysis of sliding Bloom filter (Figure~\ref{subfig:SBF}) and count distinct (Figure~\ref{subfig:Z}) is accurate.
For entropy (Figure~\ref{subfig:entropy}) the accuracy is better than anticipated indicating that our analysis here is just an upper bound, but the trend line is nearly~identical.
\subsection{Set membership on sliding windows}
We now compare SWAMP to TBF~\cite{TBF} and SWBF~\cite{slidngBloomFilterInfocom}.
Our evaluation focuses on two aspects, fixing $\varepsilon$ and changing the window size (Figure~\ref{variableW}) as well as fixing the window size and changing $\varepsilon$ (Figure~\ref{variableEps}).

As can be observed, SWAMP is considerably more space efficient than the alternatives in both cases for a wide range of window sizes and for a wide range of error probabilities.
In the tested range, it is 25-40\% smaller than the best alternative.

\subsection{Per-flow counting on sliding windows}
Next, we evaluate SWAMP for its per-flow counting functionality.
We compare SWAMP to WCSS~\cite{WCSS} that solves heavy hitters on a sliding window.
Our evaluation uses the On Arrival model, which was used to evaluate WCSS.
In that model, we perform a query for each incoming packet.
Then, we calculate the \emph{Root Mean Square Error}.
We repeated each experiment 25 times with different seeds and computed 95\% confidence intervals for SWAMP.
Note that WCSS is a deterministic algorithm and as such was run only once.

The results appear in Figure~\ref{fig:memWCSS}.
Note that the space consumption of WCSS is proportional to $\varepsilon$ and that of SWAMP to $W$.
Thus, SWAMP cannot be run for the entire range.
Yet, when it is feasible, SWAMP's error is lower on average than that of WCSS.
Additionally, in many of the configurations we are able to show statistical significance to this improvement.
Note that SWAMP becomes accurate with high probability using about $300$KB of memory while
WCSS requires about $8.3$MB to provide the same accuracy.
That is, an improvement of x27.


\subsection{Count distinct on sliding windows }
Next, we evaluate the count distinct functionality in terms of accuracy vs. space on the different datasets.
We performed $25$ runs and summarized the averaged the results.
We evaluate two functionalities: SWAMP-LB and SWAMP-MLE.
SWAMP-LB corresponds to the function \distinctLB{} in Algorithm~\ref{alg:SWAMP} and provides one sided estimation while SWAMP-MLE corresponds to the function \distinctMLE{} in Algorithm~\ref{alg:SWAMP} and provides an unbiased estimator.

Figure~\ref{fig:Z} shows the results of this evaluation.
As can be observed, SWAMP-MLE is up to x1000 more accurate than alternatives.
Additionally, SWAMP-LB also out performs the alternatives for parts of the range. Note that SWAMP-LB is the only one sided estimator in this evaluation.

\subsection{Entropy estimation on sliding window}
Figure~\ref{fig:Entropy} shows results for Entropy estimation.
As shown, SWAMP provides a very accurate entropy estimation in its entire operational range.
Moreover, our analysis in Section~\ref{anal:entropy} is conservative and SWAMP is much more accurate in practice.


	

\section{Analysis}
\label{sec:analysis}
This section aims to prove that RHHH solves the {\sc$(\delta,\epsilon,\theta)-$approximate HHH} problem (Definition~\ref{def:deltaapproxHHH}) for one and two dimensional hierarchies.
Toward that end, Section~\ref{sec:analSamples} proves the accuracy requirement while Section~\ref{anal:randHHH} proves coverage.
Section~\ref{sec:RHHH-prop} proves that RHHH solves the {\sc$(\delta,\epsilon,\theta)-$approximate HHH} problem as well as its memory and update complexity.

We model the update procedure of RHHH as a balls and bins experiment where there are $V$ bins and $N$ balls.
Prior to each packet arrival, we place the ball in a bin that is selected uniformly at random.
The first $H$ bins contain an HH update action while the next $V-H$ bins are void.
When a ball is assigned to a bin, we either update the underlying HH algorithm with a prefix obtained from the packet's headers or ignore the packet if the bin is void.
Our first goal is to derive confidence intervals around the number of balls in a bin.

\begin{definition}
We define $X^{K}_i$ to be the random variable representing the number of balls from set $K$ in bin $i$, e.g., $K$ can be all packets that share a certain prefix, or a combination of multiple prefixes with a certain characteristic.
When the set $K$ contains all packets, we use the notation $X_i$.
\end{definition}

Random variables representing the number of balls in a bin are dependent on each other.
Therefore, we cannot apply common methods to create confidence intervals.
Formally, the dependence is manifested as:\\ $\sum\nolimits_1^{V} {{X_i}}  = N.$
This means that the number of balls in a certain bin is determined by the number of balls in all other bins.

Our approach is to approximate the balls and bins experiment with the corresponding Poisson one.
That is, analyze the Poisson case and derive confidence intervals and then use Lemma~\ref{lemma:rare} to derive a (weaker) result for the original balls and bins case.

We now formally define the corresponding Poisson model.
Let $Y_1^K,...,Y_{ V }^K$ s.t. $\{Y_i^K\} \sim Poisson\left( {\frac{K}{V}} \right)$ be \textbf{independent} Poisson random variables representing the number of balls in each bin from a set of balls $K$.
That is: $\{Y_i^K\} \sim Poisson\left( {\frac{K}{V}} \right).$

\begin{lemma}[Corollary 5.11, page 103 of~\cite{Mitzenmacher:2005:PCR:1076315}]
	\label{lemma:rare}
	Let $\mathfrak E$ be an event whose probability is either monotonically increasing or decreasing with the number of balls. If $\mathfrak E$ has probability $p$ in the Poisson case then $\mathfrak E$ has probability at most $2p$ in the exact case.
\end{lemma}


\subsection{Accuracy Analysis}
\label{sec:analSamples}

We now tackle the accuracy requirement from Definition~\ref{def:deltaapproxHHH}.
That is, for every HHH prefix ($p$), we need to~prove:  $$\Pr \left( {\left| {{f_p} - \widehat {{f_p}}} \right| \le \varepsilon N} \right) \ge 1 - \delta.$$

In RHHH, there are two distinct origins of error.
Some of the error comes from fluctuations in the number of balls per bin while the approximate HH algorithm is another source of error.

We start by quantifying the balls and bins error.
Let $Y^{p}_i$ be the Poisson variable corresponding to prefix $p$.
That is, the set $p$ contains all packets that are generalized by prefix $p$.
Recall that $f_p$ is the number of packets generalized by $p$ and therefore: $E(Y^{p}_i) = \frac{f_p}{V}.$

We need to show that with probability $1-\delta_s$,  $Y^{p}_i$ is within $\epsilon_s N$ from $E(Y^{p}_i)$.
Fortunately, confidence intervals for Poisson variables are a well studied~\cite{19WaysToPoisson} and we use the method of~\cite{Wmethod} that is quoted in
Lemma~\ref{lemma:poissonConfidence}.
\begin{lemma}
	\label{lemma:poissonConfidence}
	Let $X$ be a Poisson random variable, then
	\[\Pr \left( {\left| {X - E\left( X \right)} \right| \ge {Z_{1-\delta }}\sqrt {E\left( X \right)} } \right) \le \delta,\]
where $Z_\alpha$ is the $z$ value that satisfies $\phi(z)=\alpha$ and $\phi(z)$ is the density function of the normal distribution with mean $0$ and standard deviation of $1$.
\end{lemma} Lemma~\ref{lemma:poissonConfidence}, provides us with a confidence interval for Poisson variables, and enables us to tackle the main accuracy result.
\begin{theorem}
	\label{thm:pusmain}
	If $N \ge {Z_{1 - \frac{{{\delta _s}}}{2}}}V{\varepsilon_s}^{ - 2}$ then
	\[\Pr \left( {\left| {{X_i}^pH - {f_p}} \right| \ge {\varepsilon _s}N} \right) \le {\delta _s}.\]
\end{theorem}

\begin{proof}
We use Lemma~\ref{lemma:poissonConfidence} for $\frac{\delta_s}{2}$ and get:
\[\Pr \left( \left| {{Y_i}^p - \frac{{{f_p}}}{V}} \right| \ge {Z_{1 - \frac{\delta _s}{2}}\sqrt {\frac{{{f_p}}}{V}} } \right) \le \frac{\delta_s}{2} .\]
To make this useful, we trivially bind $f_p \le N$ and get
\[\Pr \left( \left| {{Y_i}^p - \frac{{{f_p}}}{V}} \right| \ge {Z_{1 - \frac{\delta _s}{2}}\sqrt {\frac{{{N}}}{V}} } \right) \le \frac{\delta_s}{2} .\]

However, we require error of the form $\frac{\epsilon_s \cdot N}{V}$.
\[\begin{array}{l}
{\varepsilon _s}N{V^{ - 1}} \ge {Z_{1 - \frac{{{\delta _s}}}{2}}}{V^{ - 0.5}}{N^{0.5}}\\
{N^{0.5}} \ge {Z_{1 - \frac{{{\delta _s}}}{2}}}{V^{0.5}}{\varepsilon _s}^{ - 1}\\
N \ge {Z_{1 - \frac{{{\delta _s}}}{2}}}V{\varepsilon_s}^{ - 2} .
\end{array}\]
Therefore, when $N \ge {Z_{1 - \frac{{{\delta _s}}}{2}}}V{\varepsilon_s}^{ - 2}$,  we have that:
\[\Pr \left( {\left| {{Y_i}^p - \frac{{{f_p}}}{V}} \right| \ge \frac{{{\varepsilon _s}N}}{V}} \right) \le \frac{{{\delta _s}}}{2} .\]
We multiply by $V$ and get:
$$\Pr \left( {\left| {{Y_i}^pV - {f_p}} \right| \ge {\varepsilon _s}N} \right) \le \frac{{{\delta _s}}}{2} .$$
Finally, since $Y_i^{p}$ is monotonically increasing with the number of balls ($f_p$), we apply Lemma~\ref{lemma:rare} to conclude that\\
$$\Pr \left( {\left| {{X_i}^pV - {f_p}} \right| \ge {\varepsilon _s}N} \right) \le {\delta _s}.$$
\end{proof}
To reduce clutter, we denote $\NB \triangleq \NBound$.
Theorem~\ref{thm:pusmain} proves that the desired sample accuracy is achieved once $N>\NB$.

It is sometimes useful to know what happens when $N<\NB$.
For this case, we have Corollary~\ref{cor:epsN}, which is easily derived from Theorem~\ref{thm:pusmain}.
We use the notation $\varepsilon_s(N)$ to define the actual sampling error after $N$ packets.
Thus, it assures us that when $N<\NB$, $\varepsilon_s(N)>\varepsilon_s$.
It also shows that $\varepsilon_s(N)<\varepsilon_s$ when $N>\NB$.
Another application of Corollary~\ref{cor:epsN} is that given a measurement interval $N$, we can derive a value for $\varepsilon_s$ that assures correctness.
For simplicity, we continue with the notion of $\varepsilon_s$.

\begin{corollary}
	\label{cor:epsN}
	 ${\varepsilon _s}\left( N \right) \ge \sqrt {\frac{{{Z_{1 - \frac{{{\delta _s}}}{2}}}V}}{N}} .$
\end{corollary}

The error of approximate HH algorithms is proportional to the number of updates.
Therefore, our next step is to provide a bound on the number of updates of an arbitrary HH algorithm.
Given such a bound, we configure the algorithm to compensate so that the accumulated error remains within the guarantee even if the number of updates is larger than average.

\begin{corollary}
	\label{cor:oversample}
	Consider the number of updates for a certain lattice node ($X_i$).
	If $N>\NB$, then \[\Pr \left( {{X_i} \le \frac{N}{V}\left( {1 + {\varepsilon _s}} \right)} \right) \ge 1 - {\delta _s}.\]
\end{corollary}

\begin{proof}
	 We use Theorem~\ref{thm:pusmain} and get: \\
$\Pr \left( {\left| {{X_i} - \frac{N}{V}} \right| \ge {\varepsilon _s}N} \right) \le {\delta _s}.$
This implies that:\\
$\Pr \left( {{X_i} \le \frac{N}{V}\left( {1 + {\varepsilon _s}} \right)} \right) \ge 1 - {\delta _s},$
completing the proof.
\end{proof}

We explain now how to configure our algorithm to defend against situations in which a given approximate HH algorithm might get too many updates, a phenomenon we call \emph{over sample}.
Corollary~\ref{cor:oversample} bounds the probability for such an occurrence, and hence we can slightly increase the accuracy so that in the case of an over sample, we are still within the desired limit.
We use an algorithm ($\mathbb A$) that solves the {\sc {$(\varepsilon_a, \delta_a)$ - Frequency Estimation}} problem.
We define $\varepsilon_a' \triangleq \frac{\varepsilon_a}{1+\varepsilon_s}$.
According to Corollary~\ref{cor:oversample}, with probability $1-\delta_s$, the number of sampled packets is at most $(1+\varepsilon_s)\frac{N}{V}.$
By using the union bound and with probability $1-\delta_a-\delta_s$ we get:
\[\left| {{X^p} - \widehat {{X^p}}} \right| \le {\varepsilon _{a'}}\left( {1 + {\varepsilon _s}} \right)\frac{N}{V} = \frac{{{\varepsilon _a}\left( {1 + {\varepsilon _s}} \right)}}{{1 + {\varepsilon _s}}}\frac{N}{V} = {\varepsilon _a}\frac{N}{V}.\]
For example, Space Saving requires $1,000$ counters for $\epsilon_a=0.001$.
If we set $\epsilon_s = 0.001$, we now require $1001$ counters.
Hereafter, we assume that the algorithm is configured to accommodate these over samples.

\begin{theorem}
	\label{thm:PUSCombined}
	Consider an algorithm ($\mathbb{A}$) that solves the {\sc {$(\epsilon_a, \delta_a)$ - Frequency Estimation}} problem.
	If $N > \NB$, then for  $\delta \ge \delta_a + 2 \cdot \delta_s$ and $\epsilon \ge \epsilon_a + \epsilon_s$, $\mathbb{A}$ solves {\sc {$(\epsilon, \delta)$ - Frequency Estimation}}.
\end{theorem}

\begin{proof}
	As $N > \NB$, we use Theorem~\ref{thm:pusmain}.
	That is, the input solves {\sc {$(\epsilon, \delta)$ - Frequency Estimation}}.
	\begin{equation}
	\label{eq:delta2}
	\Pr \left[ {\left| {{f_p} - {X_p}V} \right| \ge {\varepsilon _s}N} \right] \le {\delta _s}.
	\end{equation}
	
	$\mathbb{A}$ solves the {\sc {$(\epsilon_a, \delta_a)$ - Frequency Estimation}} problem and provides us with an estimator $\widehat{X^p}$ that approximates $X^p$ --  the number of updates for prefix $p$.
	According to Corollary~\ref{cor:oversample}:
	\[\Pr \left( {\left| {{X^p} - \widehat {{X^p}}} \right| \le \frac{{{\varepsilon _a}N}}{V}} \right) \ge 1 - {\delta _a} - {\delta _s},\]
	and multiplying both sides by $V$ gives us:
	\begin{equation}
	\label{eq:nodelta2}
\Pr \left( {\left| {{X^p}V - \widehat {{X^p}}V} \right| \ge {\varepsilon _a}N} \right) \le {\delta _a} + {\delta _s}.
	\end{equation}
	We need to prove that: $\Pr \left( {\left| {{f_p} - \widehat {{X^p}}V} \right| \le \varepsilon N} \right) \ge 1 - \delta$.
	Recall that: $f_p = E(X^p)V$ and that $\widehat{f_p} = \widehat{X^p}V$ is the estimated frequency of $p$.
	Thus,
	\small
	\begin{align}
	&\Pr \left( {\left| {{f_p} - \widehat{f_p}} \right| \ge \varepsilon N} \right) = \Pr \left( {\left| {{f_p} - \widehat {{X^p}}V} \right| \ge \varepsilon N} \right)\notag\\
	=& \Pr \left( {\left| {{f_p} + \left( {{X^p}{V} - {X^p}{V}} \right) - {V}\widehat {{X^p}}} \right| \ge (\epsilon_a+\epsilon_s) N} \right)\label{eq:separation}
	\\ \le&\Pr \left( \left[{\left| {{f_p} - {X^p}{V}} \right| \ge {\varepsilon _s}N} \right]\vee  \left[{\left| {{X^p}{V} - \widehat {{X^p}}}{V} \right| \ge {\varepsilon _a}N}\right] \right)\notag,
	\end{align}\normalsize
	where the last inequality follows from the fact that in order for the error of~\eqref{eq:separation} to exceed $\epsilon N$, at least one of the events has to occur.
	We bound this expression using the Union bound.
\[\begin{array}{l}
\Pr \left( {\left| {{f_p} - \widehat {{f_p}}} \right| \ge \varepsilon N} \right) \le \\
\Pr \left( {\left| {{f_p} - {X^p}V} \right| \ge {\varepsilon _s}N} \right) + \Pr \left( {\left| {{X^p}V - \widehat {{X^p}}H} \right| \ge {\varepsilon _a}N} \right)  \\
\le{\delta _a} + 2{\delta _s},
\end{array}\]
where the last inequality is due to equations~\ref{eq:delta2} and~\ref{eq:nodelta2}.
\end{proof}

An immediate observation is that Theorem~\ref{thm:PUSCombined} implies accuracy, as it guarantees that with probability $1-\delta$ the estimated frequency of any prefix is within $\varepsilon N$ of the real frequency while the accuracy requirement only requires it for prefixes that are selected as HHH.

\begin{lemma}
	\label{lemma:accuracy}
	If $N > \NB$, then Algorithm~\ref{alg:Skipper} satisfies the accuracy constraint for $\delta = \delta_a+2\delta_s$ and $\epsilon = \epsilon_a+\epsilon_s$.
\end{lemma}

\begin{proof}
	The proof follows from Theorem~\ref{thm:PUSCombined}, as the frequency estimation of a prefix depends on a single HH~algorithm.
\end{proof}

\subsubsection*{Multiple Updates}
One might consider how RHHH behaves if instead of updating at most $1$ HH instance, we update $r$ independent instances. This implies that we may update the same instance more than once per packet. 
Such an extension is easy to do and still provides the required guarantees. Intuitively, this variant of the algorithm is what one would get if each packet is duplicated $r$ times. The following corollary shows that this makes RHHH converge $r$ times faster. 
\begin{corollary}
	Consider an algorithm similar to $RHHH$ with $V=H$, but for each packet we perform $r$ independent update operations. 
	If $N > \frac{\NB}{r}$, then this algorithm satisfies the accuracy constraint for $\delta = \delta_a+2\delta_s$ and $\epsilon = \epsilon_a+\epsilon_s$.
\end{corollary}
\begin{proof}
	Observe that the new algorithm is identical to running RHHH on a stream ($\mathcal{S'}$) where each packet in $\mathcal{S}$ is replaced by $r$ consecutive packets.
    Thus, Lemma~\ref{lemma:accuracy} guarantees that accuracy is achieved for $\mathcal{S'}$ after $\NB$ packets are processed.
    That is, it is achieved for the original stream ($\mathcal{S}$) after $N >\frac{\NB}{r}$ packets.
\end{proof}

\subsection{Coverage Analysis}
\label{anal:randHHH}

Our goal is to prove the coverage property of Definition~\ref{def:deltaapproxHHH}.
 That is:
$\Pr \left( \widehat {C_{q|P}} \ge C_{q|P} \right) \ge 1-\delta.$
Conditioned frequencies are calculated in a different manner for one and two dimensions.
Thus, Section~\ref{subsec:one} deals with one dimension and Section~\ref{subsec:two} with two.

We now present a common definition of the best generalized prefixes in a set.
\begin{definition}[Best generalization]
	\label{def:bestG}
	Define $G(q|P)$  as the set $\left\{ {p:p \in P,p \prec q,\neg \exists p' \in P:q \prec p' \prec p} \right\}$.
	Intuitively, $G(q|P)$ is the set of prefixes that are best generalized by $q$.
	That is, $q$ does not generalize any prefix that generalizes one of the prefixes in $G(q|P)$.
\end{definition}

\subsubsection{One Dimension}
\label{subsec:one}
We use the following lemma for bounding the error of our conditioned count estimates.
\begin{lemma}
	\label{lemma:cp}(\cite{HHHMitzenmacher})
	In one dimension, $${C_{q\mid P}} = {f_q} - \sum\nolimits_{h \in G(q|P)} {{f_h}} .$$
	\normalsize
\end{lemma}

Using Lemma~\ref{lemma:cp}, it is easier to establish that the conditioned frequency estimates calculated by Algorithm~\ref{alg:Skipper} are conservative.

\begin{lemma}
\label{lemma:sq}
The conditioned frequency estimation of Algorithm~\ref{alg:Skipper} is:
\[\widehat{C_{q|P}} = \widehat{f_q}^{+}-\sum\nolimits_{h \in G\left( {q|P} \right)} {\widehat{f_h}^- } +  2{Z_{1 - \delta }}\sqrt {N V} .\]
\end{lemma}
\begin{proof}
Looking at Line~\ref{line:cp} in Algorithm~\ref{alg:Skipper}, we get that: $$\widehat{C_{q|P}} = \widehat{f_q}^{+} + calcPred(q,P).$$
That is, we need to verify that the return value $calcPred(q,P)$ in one dimension (Algorithm~\ref{alg:randHHH}) is $\sum\nolimits_{h \in G\left( {q|P} \right)} {\widehat{f_h}^- }$.
This follows naturally from that algorithm.
Finally, the addition of $2{Z_{1 - \delta }}\sqrt {NV}$ is due to line~\ref{line:accSample}.
\end{proof}

In deterministic settings, $\widehat{f_q}^{+} -\sum\nolimits_{h \in G\left( {q|P} \right)} {\widehat{f_h}^- }$ is a conservative estimate since ${\widehat {{f_q}}^ + } \ge {f_q}$ and $f_h < \widehat{f_h}^-$.
In our case, these are only true with regard to the sampled sub-stream and the addition of $2{Z_{1 - \delta }}\sqrt {NV}$ is intended to compensate for the randomized process.


Our goal is to show that $\Pr \left(\widehat {C_{q|P}} > C_{q|P}\right) \ge 1-\delta$.
That is, the conditioned frequency estimation of Algorithm~\ref{alg:Skipper} is probabilistically conservative.



\begin{theorem}
	\label{thm:underCP}
$\Pr \left( \widehat {C_{q|P}} \ge C_{q|P} \right) \ge 1-\delta.$
\end{theorem}

\begin{proof}
	Recall that:  $$\widehat {{C_{q|P}}} = \widehat f_q^ +  - \sum\limits_{h \in G\left( {q|P} \right)} {\widehat f_h^ - + 2{Z_{1-\frac{\delta }{8}}}\sqrt {NV} }.$$

	We denote by $K$ the set of packets that may affect $\widehat {{C_{q|P}}}$.
	We split $K$ into two sets: $K^{+}$ contains the packets that may positively impact $\widehat {{C_{q|P}}}$ and $K^{-}$ contains the packets that may negatively impact it.
	
	 We use $K^{+}$ to estimate the sample error in $\widehat{ f_q}$ and $K^{-}$ to estimate the sample error in $\sum\limits_{h \in G\left( {q|P} \right)} {\widehat f_h^ -}$.
	 The positive part is easy to estimate.
	 In the negative, we do not know exactly how many bins affect the sum.
	 However, we know for sure that there are at most $N$.
	 We define the random variable $Y^K_+$ that indicates the number of balls included in the positive sum.
	  We  invoke Lemma~\ref{lemma:poissonConfidence} on $Y^{K}_+$.
	  For the negative part, the conditioned frequency is positive so $E\left(Y^K_-\right)$ is at most $\frac{N}{V}$. Hence,
	$\Pr \left( \left| {Y_K^ +  - E\left( {Y_K^ + } \right)} \right| \ge {Z_{1-\frac{\delta }{8}}}\sqrt \frac{N }{V} \right) \le \frac{\delta }{4}.$
	Similarly, we use Lemma~\ref{lemma:poissonConfidence} to bound the error of $Y_K^ -$:
	$$\Pr \left( {\left| {Y_K^ -  - E\left( {{Y_K}^ - } \right)} \right| \ge {Z_{1-\frac{\delta }{8}}}\sqrt  \frac{N}{V} } \right) \le \frac{\delta }{4}.$$\\
	$Y^{K}_+$ is monotonically increasing with any ball and $Y_{K}^-$ is monotonically decreasing with any ball.
	Therefore,  we can apply Lemma~\ref{lemma:rare} on each of them and conclude:
\[\begin{array}{l}
\Pr \left( {\widehat {{C_{q|P}}} \ge {C_{q|P}}} \right)\le\\
 2\Pr \left( {H\left( {Y_K^ -  + Y_K^ + } \right)   \ge VE\left( {Y_K^ -  + Y_K^ + } \right) + 2{Z_{1 - \frac{\delta }{8}}}\sqrt {NV} } \right)\\
\le 1-2\frac{\delta }{2} = 1-\delta.
\end{array}\]
\end{proof}
\normalsize
\begin{theorem}
	If $N > \NB$, Algorithm~\ref{alg:Skipper} solves the {\sc$(\delta, \varepsilon, \theta)$ - Approximate HHH} problem for $\delta = \delta_a + 2\delta_s$ and $\varepsilon = \varepsilon_s + \varepsilon_a$.
\end{theorem}
\begin{proof}
We need to show that the accuracy and coverage guarantees hold.
Accuracy follows from Lemma~\ref{lemma:accuracy} and coverage follows from Theorem~\ref{thm:underCP} that implies that for every non heavy hitter prefix (q),  $\widehat{C_{q|P}}<\theta N$ and thus:  $$\Pr \left( {{C_{q|P}} < \theta N} \right) \ge 1 - \delta.$$
\end{proof}


\section{Discussion}
\label{sec:discussion}

In modern networks, operators are likely to require multiple measurement types. To that end, this work suggests SWAMP, a unified algorithm that monitors four common measurement metrics in constant time and compact space. Specifically, SWAMP approximates the following metrics on a sliding window: Bloom filters, per-flow counting, count distinct and entropy estimation.
For all problems, we proved formal accuracy guarantees and demonstrated them on real Internet traces.

Despite being a general algorithm, SWAMP advances the state of the art for all these problems.
For sliding Bloom filters, we showed that SWAMP is memory succinct for constant false positive rates and that it reduces the required space by 25\%-40\% compared to previous approaches~\cite{slidngBloomFilterInfocom}.
In per-flow counting, our algorithm outperforms WCSS~\cite{WCSS} -- a state of the art window algorithm.
When compared with $1+\varepsilon$ approximation count distinct algorithms~\cite{SlidingHLL,Fusy-HLL}, SWAMP asymptotically improves the query time from $O(\varepsilon^{-2})$ to a constant. It is also up to x1000 times more accurate on real packet traces. For the entropy estimation on a sliding window~\cite{SlidingEntropy}, SWAMP reduces the update time to a constant.

While SWAMP benefits from the compactness of TinyTable~\cite{TinyTable}, most of its space reductions inherently come from using fingerprints rather than sketches. For example, all existing count distinct and entropy algorithms require $\Omega(\epsilon^{-2})$ space for computing a $1+\epsilon$ approximation. SWAMP can compute the \emph{exact} answers using $O(W\log W)$ bits. Thus, for a small $\epsilon$ value, we get an asymptotic reduction by storing the fingerprints on \emph{any} compact table.



Finally, while the formal analysis of SWAMP is mathematically involved, the actual code is short and simple to implement. This facilitates its adoption in network devices and SDN. In particular, OpenBox~\cite{openbox} demonstrated that sharing common measurement results across multiple network functionalities is feasible and efficient. Our work fits into~this~trend.

\bibliographystyle{abbrv}

\bibliography{refs}

\end{document}